\documentclass[prd,onecolumn,nofootinbib,amsfonts,amssymb,amsmath,12pt,preprint]{revtex4}
\usepackage{latexsym,graphicx,psfrag,here,bm,amsmath}
\newcommand{\nn}{\nonumber}
\newcommand{\Eref}[1]{Eq.~(\ref{#1})}
\newcommand{\Sref}[1]{Section~\ref{#1}}
\newcommand{\Fref}[1]{Fig.~\ref{#1}}

\newcommand{\bdll}{b\to d\,\ell^{+}\ell^{-}}

\newcommand{\bsll}{b\to s\,\ell^{+}\ell^{-}}

\newcommand{\IR}{{\text{IR}}}

\newcommand{\wtC}{\widetilde{C}}
\newcommand{\wtO}{\widetilde{O}}
\newcommand{\mubeps}{\left[ \frac{\mu}{m_b} \right]^{2\epsilon}}

\newcommand{\pole}{\text{pole}}
\newcommand{\st}{\rule[-3mm]{0mm}{9mm}}
\newcommand{\Brems}{\text{Brems}}

\newcommand{\ibid}[3]{{ibid.} {\bf #1}  (#2) #3}
\newcommand{\zpc}[3]{{Z. Phys. C} {\bf #1}  (#2) #3}

\newcommand{\plb}[3]{{Phys. Lett.} {\bf B #1}  (#2) #3}
\newcommand{\prdd}[3]{{Phys. Rev.} {\bf D #1}  (#2) #3}
\renewcommand{\prl}[3]{{Phys. Rev. Lett.} {\bf #1}  (#2) #3}
\newcommand{\npb}[3]{{Nucl. Phys.} {\bf B #1}  (#2) #3}
\newfont{\typew}{ectt12}
\newcommand{\hepph}[1]{{\tt{hep-ph/#1}}}
\newcommand{\hepth}[1]{{\tt{hep-th/#1}}}
\newcommand{\hepex}[1]{{\tt{hep-ex/#1}}}

\newcommand{\pd}{\partial}

\newcommand{\eff}{\text{eff}}

\newcommand{\ordo}[1]{\ensuremath{{\cal O}(#1)}}

\renewcommand{\Re}{\text{Re}}

\newcommand{\alphas}{\ensuremath{\alpha_{\text{s}}}}

\newcommand{\alphasFourPi}{\ensuremath{\frac{\alphas}{4 \pi}}}

\newcommand{\s}{\hat{s}}

\newcommand{\bea}{\begin{eqnarray}}
\newcommand{\eea}{\end{eqnarray}}
\newcommand{\beq}{\begin{equation}}
\newcommand{\eeq}{\end{equation}}

\def\bra{\langle}
\def\ket{\rangle}

\def\qr{q \, r}

\def\a{\alpha}
\def\b{\beta}
\def\g{\gamma}
\def\d{\delta}
\def\e{\epsilon}
\def\p{\pi}

\def\l{\lambda}
\def\m{\mu}
\def\n{\nu}

\begin{document}

%
%

\title{Virtual- and bremsstrahlung corrections
       to $\bm{b \rightarrow d \,\ell^+\ell^-}$ in the standard model }
\author{H.~M.~Asatrian}
\email{hrachia@jerewan1.yerphi.am}
\affiliation{Yerevan Physics Institute, 2 Alikhanyan Br., 375036
    Yerevan, Armenia}
\author{K.~Bieri}
\email{bierik@itp.unibe.ch}
\author{C.~Greub}
\email{greub@itp.unibe.ch}
\author{M.~Walker}
\email{walker@itp.unibe.ch,mwalker@bioc.unizh.ch}
\affiliation{Institut f\"ur Theoretische Physik, Universit\"at Bern,
  CH-3012 Bern, Switzerland}
\begin{abstract}
We present the calculation of the virtual- and bremsstrahlung corrections of
$\ordo{\alpha_s}$ to the matrix elements $\bra d\,\ell^+ \ell^-|O_i|b \ket$.
This is the missing piece in the full next-to-next-to-leading logarithmic
(NNLL) results for various observables associated with the process $B \to X_d
\ell^+ \ell^-$, like the branching ratio, the CP-rate asymmetry and the
forward-backward asymmetry.  This paper is an extension of analogous
calculations done by some of us for the process $B \to X_s \ell^+ \ell^-$. As
the contributions of the diagrams induced by the four-quark operators $O_1^u$
and $O_2^u$ with a $u$-quark running in the quark loop are strongly CKM
suppressed, they were omitted in the analysis of $B \to X_s \ell^+ \ell^-$.
This is no longer possible for $B \to X_d \ell^+ \ell^-$, as the corresponding
contributions are not suppressed. The main new work therefore consists of
calculating the ${\cal O}(\alpha_s)$ corrections to $\bra d \,\ell^+
\ell^-|O_{1,2}^u|b \ket$. In this paper we restrict ourselves to the range
$0.05 \le s/m_b^2 \le 0.25$ ($s$ is the invariant mass of the lepton pair),
which lies above the $\rho$- and $\omega$-resonances and below the
$J/\psi$-resonance. We present the analytic results for the mentioned
observables related to the process $B \to X_d \ell^+ \ell^- $ as expansions in
the small parameters $\s = s/m_b^2$, $z = m_c^2/m_b^2$ and $s/(4\, m_c^2)$. In
the phenomenological analysis at the end of the paper we discuss the impact of
the NNLL corrections on the observables mentioned above.
\end{abstract}
\maketitle

%
%
\section{Introduction}
It is well-known that various observables associated with inclusive rare
$B$-decays like $B \to X_{s,d} \gamma$ and $B \to X_{s,d} \ell^+ \ell^-$
sensitively depend on potential new physics contributions. But even in the
absence of new physics these observables are important, because they provide
checks on the one-loop structure of the Standard Model (SM) theory and 
can be used to gain information on the Cabibbo-Kobayashi-Maskawa (CKM) matrix
elements $V_{ts}$ and $V_{td}$, which are difficult to measure directly.

At present, a lot of data already exists on ${\rm BR}(B \to X_s \gamma)$
\cite{CLEOold, CLEOmed, CLEOnew, BRALEPH, BELLE01, BABAR02, Jessop2003} and on
${\rm BR}(B \to X_s \ell^+ \ell^-)$ \cite{BELLEbsll2, BELLEbsll3, Babarbsll}
and it is expected that in the future also data on the CKM suppressed
counterparts, i.e. on ${\rm BR}(B \to X_d \gamma)$ and on ${\rm BR}(B \to X_d
\ell^+ \ell^-)$ will become available. The same holds for experimental
information on additional observables, like CP-rate asymmetries or
forward-backward asymmetries in the decays $B \to X_{s,d} \ell^+ \ell^-$.

In order to fully exploit and interpret the experimental data, it is obvious
that precise calculations in the SM (or certain extensions thereof) are
needed.  The main problem in the theoretical description of the decay $B \to
X_s \ell^+ \ell^-$ is due to the long-distance contributions induced by
$\bar{c}c$ resonant states and in principle also by $\bar{u}u$ resonant
states.  The latter are, however, strongly CKM suppressed. This suppression is
not present in the case of $B \to X_d \ell^+ \ell^-$, as the CKM factors
involved in the contributions from $\bar{c}c$ and $\bar{u}u$ resonant states
are of the same order.  When the invariant mass $\sqrt{s}$ of the lepton pair
is close to the mass of a resonance, only model dependent predictions for such
long distance contributions are available at present. It is therefore unclear
whether the theoretical uncertainty can be reduced to less than $\pm 20\%$
when integrating over these domains \cite{Ligeti:1996yz}. 

However, restricting $\sqrt{s}$ to a region below the $\bar{c}c$ resonances,
the long distance effects in $B \to X_s \ell^+ \ell^-$ are under control.  The
same is true for $B \to X_d \ell^+ \ell^-$ when choosing a region of
$\sqrt{s}$ which is below the $J/\psi$- and above the $\rho,\omega$-resonance
regions. It turns out that in those ranges of $\sqrt{s}$ the corrections to
the pure perturbative picture can be analyzed within the heavy quark effective
theory (HQET). In particular, all available studies indicate that for the
region $0.05<\hat{s}=s/m_b^2<0.25$ the non-perturbative effects are below
10$\%$ \cite{Falk:1994dh, Ali:1997bm, chen:1997, Buchalla:1998ky,
Buchalla:1998mt, Kruger:1996dt}.  Consequently, observables like differential
decay rates, forward-backward asymmetries and CP-rate asymmetries for $B \to
X_{s,d} \ell^+ \ell^-$ can be precisely predicted in this region of $\sqrt{s}$
using renormalization group improved perturbation theory.  It was pointed out
in the literature that the differential decay rate and the forward-backward
asymmetry in $B \to X_s \ell^+ \ell^-$ are particularly sensitive to new
physics in this kinematical window \cite{Ball,Lunghi,Silvestrini}.

In the context of the SM there exist computations of next-to-leading
logarithmic (NLL) QCD corrections to the branching ratios for $B \to X_s
\gamma$ \cite{Mikolaj, counterterm, Adel, GH, INFRARED, AG91, Pott, GHW} and
$B\to X_d \gamma$ and the corresponding CP-rate asymmetries \cite{Ali:1998rr,
Hurth:2001ja, Hurth:2001yb}. Next-to-next-to-leading logarithmic (NNLL) QCD
corrections to the branching ratio \cite{Bobeth:2000, Gambino, AAGW, Asa2,
Ghinculov:2003bx} and the forward-backward asymmetry in $B \to X_s \ell^+
\ell^-$ are also available \cite{Adrian1, Asatrian:2002va,
Adrian2,Asatrian:2003yk}.  For a recent review see e.g.~\cite{hurth_review}.

The corresponding NNLL results for the process $B \to X_d \ell^+ \ell^-$ are
missing, however. The aim of the present paper is to close this gap.  The main
difference between the calculations for $B \to X_s \ell^+ \ell^-$ and $B \to
X_d \ell^+ \ell^-$ lies in the contributions of the current-current
operators. In the existing NNLL calculations of $B \to X_s \ell^+ \ell^-$ only
those associated with $O_1^c$ and $O_2^c$ were included at the two-loop level
because those induced by $O_1^u$ and $O_2^u$ are strongly CKM suppressed (see
\Sref{section:eff} for the definition of the operators $O_{1,2}^{u,c}$).  For
$B \to X_d \ell^+ \ell^-$ the contributions generated by $O_1^u$ and $O_2^u$
are no longer CKM suppressed and have to be taken into account as well.  At
first sight, it seems that the two-loop matrix elements of $O_1^u$ and $O_2^u$
can be straightforwardly obtained from those of $O_1^c$ and $O_2^c$ by simply
taking the limit $m_c \to 0$.  This is, however, not possible for some of the
diagrams in \Fref{fig:O1O2}, because the two-loop matrix elements of $O_1^c$
and $O_2^c$ were derived by doing various expansions. In particular, one of
the expansion parameters is $s/(4 m_c^2)$, which is formally $\ll1$ when
restricting $\sqrt{s}$ to the window discussed above. Obviously, the analogous
quantity for the $u$-quark contribution, $s/(4 m_u^2)$, cannot be used as an
expansion parameter, which implies that genuinely new calculations for the
$u$-quark contributions are needed. As discussed in \Sref{section:O1O2}, the
calculations of certain diagrams associated with $O_{1,2}^u$ are even more
involved than those associated with $O_{1,2}^c$.  To derive the new results,
we used dimension-shifting techniques in order to reduce certain tensor
integrals to scalar ones and integration-by-parts techniques to further
simplify the scalar integrals \cite{Tarasov, baikov}.
  
As the main emphasis of this paper is the derivation of the matrix elements
$\bra d\,\ell^+ \ell^- |O_{1,2}^u|b \ket$ at order $\alpha_s$, we keep the
phenomenological analysis relatively short.  In particular, we do not take
into account power corrections, but merely illustrate how the NNLL
contributions modify the scale dependences of the branching ratio, the
forward-backward asymmetry and the CP-rate asymmetry. A more detailed
phenomenology, including power corrections, will be presented elsewhere.

The paper is organized as follows: In \Sref{section:eff} we present the
effective Hamiltonian for the decay $b\to
X_d\ell^+\ell^-$. \Sref{section:O1O2} is devoted to the virtual $\ordo{\a_s}$
corrections to the operators $O_1^{u,c}$ and $O_2^{u,c}$. Subsequently,
\Sref{Section:O789} presents the corresponding contributions to the form
factors of the operators $O_7,\,O_8,\,O_9$ and $O_{10}$. With these results at
hand, we discuss in \Sref{section:decay_width} the corrections to the decay
width of $B\to X_d\ell^+\ell^-$. In \Sref{section:phenom} we show some
applications of our results. A summary of the paper is presented in
\Sref{section:summary}.  The appendices contain technical details about the
performed calculation: Appendix~\ref{section:techniques} explains the
dimension-shifting and integration-by-parts techniques. These techniques are
then applied to the calculation of diagrams~\ref{fig:O1O2}d), which is
presented in Appendix~\ref{section:diagram_d}. Appendix~\ref{appendix:wilson}
outlines a procedure on how to calculate the evolution matrix for the Wilson
coefficients as a power series in $\alpha_s$. Appendix~\ref{appendix:oneloop}
contains one-loop matrix elements needed in the calculation of the
counterterms. Finally, in Appendix~\ref{appendix:brems} we present the results
for those bremsstrahlung contributions which are free of infrared and
collinear divergences.

%
%
\section{Effective Hamiltonian}
\label{section:eff}
The appropriate framework for studying QCD corrections to rare $B$-decays in a
systematic way is the effective Hamiltonian technique. For the specific decay
channels $B\rightarrow X_s \ell^+\ell^-$ and $B \to X_d \ell^+ \ell^-$
($\ell=\mu$, $e$), the effective Hamiltonian is derived by integrating out the
$t$-quark, the $W$-boson and the $Z^0$-boson.  In the process $B \to X_s
\ell^+ \ell^-$, the appearing CKM combinations are $\lambda_u$, $\lambda_c$
and $\lambda_t$, where $\lambda_i=V_{ib}V_{is}^*$. Since $|\lambda_u|$ is much
smaller than $|\lambda_c|$ and $|\lambda_t|$, it is a safe approximation to
set $\lambda_u$ equal to zero. Using then the unitarity properties of the CKM
matrix, the CKM dependence of the Hamiltonian can be written as a global
factor $\lambda_t$. In the case of $B \to X_d \ell^+ \ell^-$, all three
quantities $\xi_i = V_{ib} V_{id}^*$ ($i=u,c,t$) are of the same order of
magnitude. Therefore, as no approximation is possible, the CKM dependence does
not globally factorize. The effective Hamiltonian reads
\begin{equation}
    {\cal H}_{\eff} = \frac{4\,G_F}{\sqrt{2}} \left[ \sum_{i=1}^{2} C_i
        (\xi_c\,O_i^c + \xi_u O_i^u) - \xi_t \sum_{i=3}^{10} C_i
        \, O_i \right] \, .
\end{equation}
We choose the operator basis according to \cite{Bobeth:2000}:
\begin{equation}
\label{oper}
\begin{array}{rclrcl}
    O_1^u  & = & (\bar{d}_{L}\gamma_{\mu} T^a u_{L })
                 (\bar{u}_{L }\gamma^{\mu} T^a b_{L}), &
    O_2^u  & = & (\bar{d}_{L}\gamma_{\mu}  u_{L })
                 (\bar{u}_{L }\gamma^{\mu} b_{L}), \vspace{0.2cm} 
                 \\ \vspace{0.3cm}
    O_1^c  & = & (\bar{d}_{L}\gamma_{\mu} T^a c_{L })
                 (\bar{c}_{L }\gamma^{\mu} T^a b_{L}), &
    O_2^c  & = & (\bar{d}_{L}\gamma_{\mu}  c_{L })
                 (\bar{c}_{L }\gamma^{\mu} b_{L}), \\ \vspace{0.2cm}
    O_3    & = & (\bar{d}_{L}\gamma_{\mu}  b_{L })
                 \sum_q (\bar{q}\gamma^{\mu}  q), &
    O_4    & = & (\bar{d}_{L}\gamma_{\mu} T^a b_{L })
                 \sum_q (\bar{q}\gamma^{\mu} T^a q), \\ \vspace{0.2cm}
    O_5    & = & (\bar{d}_L \gamma_{\mu} \gamma_{\nu} \gamma_{\rho}b_L)
                 \sum_q(\bar{q} \gamma^{\mu} \gamma^{\nu}\gamma^{\rho}q), &
    O_6    & = & (\bar{d}_L \gamma_{\mu} \gamma_{\nu} \gamma_{\rho} T^a b_L)
                 \sum_q(\bar{q} \gamma^{\mu} \gamma^{\nu} \gamma^{\rho} T^a q),
                 \\ \vspace{0.2cm}
    O_7    & = & \frac{e}{g_s^2} m_b (\bar{d}_{L} \sigma^{\mu\nu}
                 b_{R}) F_{\mu\nu}, &
    O_8    & = & \frac{1}{g_s} m_b (\bar{d}_{L} \sigma^{\mu\nu}
                 T^a b_{R}) G_{\mu\nu}^a, \\ \vspace{0.2cm}
    O_9    & = & \frac{e^2}{g_s^2}(\bar{d}_L\gamma_{\mu} b_L)
                 \sum_\ell(\bar{\ell}\gamma^{\mu}\ell), &
    O_{10} & = & \frac{e^2}{g_s^2}(\bar{d}_L\gamma_{\mu} b_L)
                 \sum_\ell(\bar{\ell}\gamma^{\mu} \gamma_{5} \ell),
\end{array}
\end{equation}
where the subscripts $L$ and $R$ refer to left- and right-handed 
components of the fermion fields, respectively.

The factors $1/g_s^2$ in the definition of the operators $O_7$, $O_9$ and
$O_{10}$ as well as the factor $1/g_s$ present in $O_8$ have been chosen by
Misiak~\cite{Misiak:1993bc} in order to simplify the organization of the
calculation. With these definitions, the one-loop anomalous dimensions [needed
for a leading logarithmic (LL) calculation] of the operators $O_i$ are all
proportional to $g_s^2$, while two-loop anomalous dimensions [needed for a
next-to-leading logarithmic (NLL) calculation] are proportional to $g_s^4$,
etc.

After this important remark we now outline the principal steps which lead to a
LL, NLL, and a NNLL prediction for the decay amplitude for 
$b \to d\,\ell^+ \ell^-$:
\begin{enumerate}
\item A matching calculation between the full SM theory and the effective
    theory has to be performed in order to determine the Wilson coefficients
    $C_i$ at the high scale $\mu_W\sim m_W,m_t$. At this scale, the
    coefficients can be worked out in fixed order perturbation theory,
    i.e. they can be expanded in $g_s^2$:
    \begin{equation}
        C_i(\mu_W) = C_i^{(0)}(\mu_W) + \frac{g_s^2}{16\pi^2} C_i^{(1)}(\mu_W)
        + \frac{g_s^4}{(16\pi^2)^2} C_i^{(2)}(\mu_W) + {\cal O}(g_s^6) \, .
    \end{equation}
At LL order, only $C_i^{(0)}$ are needed, at NLL order also $C_i^{(1)}$,
etc. The coefficient $C_7^{(2)}$ was worked out in
Refs.~\cite{Adel,GH,INFRARED}, while $C_{9}^{(2)}$ and $C_{10}^{(2)}$ were
calculated in Ref.~\cite{Bobeth:2000}.
\item The renormalization group equation (RGE) has to be solved in order to
get the Wilson coefficients at the low scale $\mu_b \sim m_b$. For this RGE
step the anomalous dimension matrix $\gamma(\alphas)$, which can be expanded
as
\begin{equation}
  \gamma(\alphas) = \gamma^{(0)} \alphasFourPi + \gamma^{(1)}
  \left(\alphasFourPi \right)^2 + \gamma^{(2)} \left( \alphasFourPi \right)^3
  + \ldots\, ,
\end{equation}
is required up to the term proportional to $\gamma^{(2)}$ when aiming at a NNLL
calculation.  After the matching step and the RGE evolution, the Wilson
coefficients $C_i(\mu_b)$ can be decomposed into a LL, NLL and NNLL part
according to
    \begin{equation}
        \label{wilsondecomplow}
        C_i(\mu_b) = C_i^{(0)}(\mu_b) + \frac{g_s^2(\mu_b)}{16\pi^2}
C_i^{(1)}(\mu_b) + \frac{g_s^4(\mu_b)}{(16\pi^2)^2} C_i^{(2)}(\mu_b) + {\cal
O}(g_s^6) \, .
    \end{equation}
We stress at this point that the entries in $\gamma^{(2)}$ which describe the
three-loop mixings of the four-quark operators $O_1 - O_6$ into the operator
$O_9$ have been calculated only recently \cite{Gambino}. In order to include
the impact of these new ingredients on the Wilson coefficient $C_9(\mu_b)$, we
had to reanalyze the RGE step. In Appendix \ref{appendix:wilson}, we derive a
practical formula for the evolution matrix $U(\mu_b,\mu_W)$ at NNLL order,
generalizing existing formulas at NLL order (see e.g.~\cite{Buchalla:1995vs}).
\item In order to get the decay amplitude, the matrix elements $\bra d\,\ell^+
    \ell^-|O_i(\mu_b)|b \ket$ have to be calculated. At LL precision, only the
    operator $O_9$ contributes, as this operator is the only one which at the
    same time has a Wilson coefficient starting at lowest order and an
    explicit $1/g_s^2$ factor in the definition. Hence, at NLL precision, QCD
    corrections (virtual and bremsstrahlung) to the matrix element of $O_9$
    are needed. They have been calculated in
    Refs.~\cite{Misiak:1993bc,Buras:1994dj}. At NLL precision, also the other
    operators start contributing, viz. $O_7(\mu_b)$ and $O_{10}(\mu_b)$
    contribute at tree-level and the four-quark operators $O_1,...,O_6$ at
    one-loop level. Accordingly, QCD corrections to the latter matrix elements
    are needed for a NNLL prediction of the decay amplitude.
\end{enumerate}

The formally leading term $\sim (1/g_s^2) C_9^{(0)}(\mu_b)$ to the amplitude
for $b \to d\,\ell^+ \ell^-$ is smaller than the NLL term $\sim (1/g_s^2)
[g_s^2/(16 \pi^2)] \, C_9^{(1)}(\mu_b)$ \cite{Grinstein:1989}. We adapt our
systematics to the numerical situation and treat the sum of these two terms as
a NLL contribution. This is, admittedly some abuse of language, because the
decay amplitude then starts out with a term which is called NLL.

As pointed out in step 3), ${\cal O}(\alpha_s)$ QCD corrections to the matrix
elements $\bra d \,\ell^+ \ell^-|O_i(\mu_b)|b \ket$ have to be calculated in
order to obtain the NNLL prediction for the decay amplitude. In the present
paper we systematically evaluate virtual corrections of order $\alpha_s$ to
the matrix elements of $O_1$, $O_2$, $O_7$, $O_8$, $O_9$ and $O_{10}$.  As the
Wilson coefficients of the gluonic penguin operators $O_3,...,O_6$ are much
smaller than those of $O_1$ and $O_2$, we neglect QCD corrections to their
matrix elements. 
We also systematically include gluon bremsstrahlung
corrections to the matrix elements of the operators just mentioned. Some of
these contributions contain infrared and collinear singularities, which are
canceled when combined with the virtual corrections.

%
%
\section{Virtual $\bm{\ordo{\alpha_s}}$ Corrections to the matrix elements
  $\bm{\bra d \,\ell^+ \ell^-| O_{1,2}^{u,c}|b \ket}$}
\label{section:O1O2}
\begin{figure}[t]
    \begin{center}
    \includegraphics[width=\textwidth]{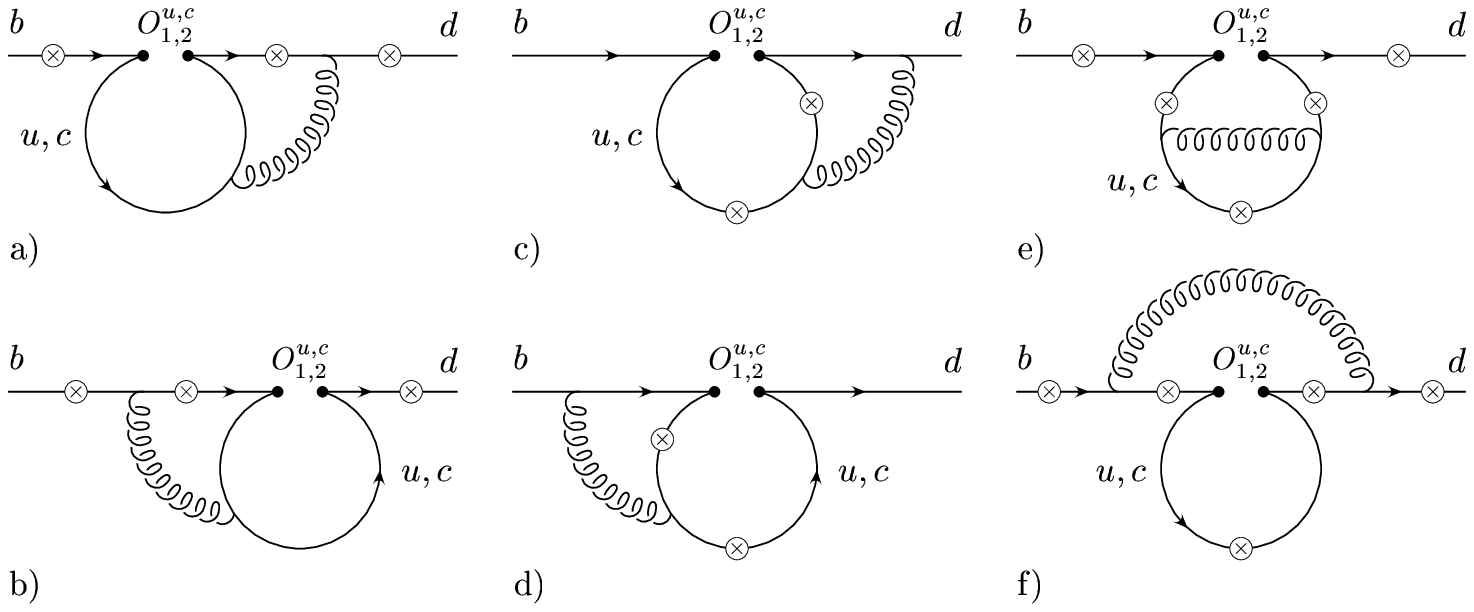}
    \caption{Complete list of two-loop Feynman diagrams for $b \to
      d\,\gamma^*$ associated with the operators $O_1^{u,c}$ and
    $O_2^{u,c}$. The fermions ($b$-, $d$-, $u$- and $c$-quarks) are represented
    by solid lines, whereas the curly lines represent gluons. The
    circle-crosses denote the possible locations where the virtual photon
    (which then splits into a lepton pair) is emitted.}
    \label{fig:O1O2}
    \end{center}
\end{figure}
In this section we present the calculation of the virtual $\ordo{\alpha_s}$
corrections to the matrix elements of the current-current operators
$O_1^{u,c}$ and $O_2^{u,c}$. Using the naive dimensional regularization scheme
(NDR) in $d=4-2\,\epsilon$ dimensions, both ultraviolet and infrared
singularities show up as $1/\epsilon^n$ poles ($n=1,2$).  The ultraviolet
singularities cancel after including the counterterms. Collinear singularities
are regularized by retaining a finite down quark mass $m_d$. They are
canceled together with the infrared singularities at the level of the decay
width when taking the bremsstrahlung process $\bdll g$ into account.  We use
the $\overline{\text{MS}}$ renormalization scheme, i.e. we introduce the
renormalization scale in the form $\overline{\mu}^{\,2} = \mu^2
\exp(\gamma_E)/(4\,\pi)$ followed by minimal subtraction. The precise
definition of the evanescent operators, which is necessary to fully specify
the renormalization scheme, will be given later.

Gauge invariance implies that the QCD corrected matrix elements of the
operators $O_i^q$ can be written as
\begin{equation}
    \label{formdef}
    \bra d\,\ell^+\ell^-|O_i^q|b\ket = \hat{F}_{i,q}^{(9)} \bra O_9
        \ket_{\text{tree}} + \hat{F}_{i,q}^{(7)} \bra O_7 \ket_{\text{tree}}
        \quad  (i=1,2;\,\,q=u,c)\,,
\end{equation}
where $\bra O_9 \ket_{\text{tree}}$ and $\bra O_7 \ket_{\text{tree}}$ are the
tree-level matrix elements of $O_9$ and $O_7$, respectively. Equivalently, we
may write
\begin{equation}
    \label{formdeftilde}
    \bra d\,\ell^+\ell^-|O_i^q|b\ket = - \frac{\alpha_s}{4\,\pi} \, \left[
    F_{i,q}^{(9)} \bra \widetilde{O}_9 \ket_{\text{tree}} + F_{i,q}^{(7)} \bra
    \widetilde{O}_7 \ket_{\text{tree}} \right],
\end{equation}
where the operators $\widetilde{O}_7$ and $\widetilde{O}_9$ are defined as
\begin{equation}
\label{Otilde}
    \widetilde{O}_7 = \frac{\alpha_s}{4\, \pi} \, O_7 , \quad \widetilde{O}_9
    = \frac{\alpha_s}{4\, \pi} \, O_9 .
\end{equation}
We present the final results for the QCD corrected matrix elements in the form
of \Eref{formdeftilde}.
The full set of the diagrams contributing at $\ordo{\alpha_s}$ to the matrix
elements
\begin{equation}
    M_i^q = \bra d\, \ell^+ \ell^- |O_i^q| b \ket
\end{equation}
is shown in Fig.~\ref{fig:O1O2}. As indicated, the diagrams associated
with $O_1^{u,c}$ and $O_2^{u,c}$ are topologically identical. They differ only
in the color structure. While the matrix elements of the operator $O_2^{u,c}$
all involve the color structure
\begin{equation*}
    \sum_a T^a T^a = C_F {\bf 1}, \quad C_F = \frac{N_c^2-1}{2\, N_c}\, ,
\end{equation*}
there are two possible color structures for the corresponding diagrams of
$O_1^{u,c}$, viz
\begin{equation*}
    \tau_1 = \sum_{a,b} T^a T^b T^a T^b \quad\text{and} \quad \tau_2 =
    \sum_{a,b} T^a T^b T^b T^a.
\end{equation*}
The structure $\tau_1$ appears in diagrams \ref{fig:O1O2}a)-d), and
$\tau_2$ enters diagrams \ref{fig:O1O2}e) and
\ref{fig:O1O2}f). Using the relation
\[
    \sum_a \, T^a_{\alpha\beta} T^a_{\gamma\delta} = -\frac{1}{2\,
        N_c}\,\delta_{\alpha\beta} \delta_{\gamma\delta} + \frac{1}{2}\,
        \delta_{\alpha\delta} \delta_{\beta\gamma} ,
\]
we find that $\tau_1=C_{\tau_1}{\bf 1}$ and $\tau_2=C_{\tau_2}{\bf 1}$, with
\[
    C_{\tau_1} = - \frac{N_c^2-1}{4\, N_c^2} \quad\text{and}\quad C_{\tau_2} =
    \frac{\left(N_c^2-1\right)^2}{4\, N_c^2} \, .
\]
Inserting $N_c=3$, the color factors are $C_F=\frac{4}{3},
C_{\tau_1}=-\frac{2}{9}$ and $C_{\tau_2}=\frac{16}{9}$. The contributions from
$O_1^{u,c}$ are obtained by multiplying those from $O_2^{u,c}$ by the
appropriate factors, i.e. by $C_{\tau_1} / C_F = -\frac{1}{6}$ and $C_{\tau_2}
/ C_F = \frac{4}{3}$, respectively.  As the renormalized $\ordo{\alpha_s}$
contributions of the operators $O_1^c$ and $O_2^c$ are discussed in detail in
Ref.~\cite{AAGW}, we only discuss the calculations of the contributions from
$O_1^u$ and $O_2^u$ to the individual form factors.

The rest of this section is organized as follows: We discuss the calculations
of the diagrams~\ref{fig:O1O2}a)-e) for the operators $O_{1,2}^u$.  Notice
that all results are given as an expansion in the small quantity $\s=s/m_b^2$,
where $s$ is the invariant mass squared of the lepton pair, and that we keep
only terms up to $\ordo{\s^3}$. After deriving the counterterms that cancel
the divergences of the diagrams mentioned above, we present the renormalized
contributions to the form factors. We postpone the discussion of
diagrams~\ref{fig:O1O2}f) as it turns out to be more convenient to take them
into account when discussing the virtual corrections to $O_9$.

%
%
\subsection{Diagrams \ref{fig:O1O2}a) and b)}
The calculation of the contributions to $F_{2,u}^{(7)}$ and $F_{2,u}^{(9)}$
from the diagrams in Figs.~\ref{fig:O1O2}a) and \ref{fig:O1O2}b) opposes no
difficulties, as it can be performed by using the Mellin-Barnes approach
\cite{Smirnov}.  Alternatively, one may get the results directly from the
corresponding form factors of the $\bsll$ transition by taking the limit $m_c
\to 0$. The form factors associated with the diagrams in Fig.~\ref{fig:O1O2}a)
are given by
\begin{align}
    F_{2,u}^{(9)}[a] = C_F \cdot \Bigg[ & -\frac{2}{27\,\epsilon^2} +
        \left(\frac{1}{\epsilon} + 4\,L_\mu \right) \left(-\frac{19}{81} +
        \frac{4}{27}\,L_s - \frac{4}{27}\,i\pi \right) -
        \frac{8}{27\,\epsilon}\,L_\mu \nn \\ &- \frac{16}{27}\,L_\mu^2
        + \left( -\frac{463}{486} - \frac{38\,i\pi}{81} + \frac{5\,\pi^2}{27}
        \right) - \frac{4}{27}\,\s + \left( -\frac{1}{27} - \frac{2}{27}\, L_s
        \right) \s^2 \nn \\ & + \left(-\frac{4}{243}-\frac{8}{81}\, L_s
        \right) \s^3 + \frac{26}{81}\, L_s + \frac{8}{27}\, i \pi\,L_s -
        \frac{2}{27}\, L_s^2 \Bigg], \\ \nn \\ F_{2,u}^{(7)}[a] = C_F
        \cdot \Bigg[ & \frac{1}{27}\left(\frac{1}{\epsilon} + 4\,L_\mu \right)
        +\frac{37}{162} +\frac{2}{27}\,i\pi + \frac{2}{27}\,\s \left(
        1+\s+\s^2 \right) L_s \Bigg],\nn
\end{align}
where
\[
    L_s = \ln(\s) \quad\text{and}\quad L_\mu =
    \ln\!\left(\!\frac{\mu}{m_b}\right) \, .
\]
For the sum of the diagrams in
Fig.~\ref{fig:O1O2}b) we find
\begin{align}
    F_{2,u}^{(9)}[b] = C_F \cdot \Bigg[& -\frac{2}{27\,\epsilon^2} +
        \left(\frac{1}{\epsilon} + 4\,L_\mu \right) \left( \frac{1}{81} -
        \frac{4}{135}\,\s - \frac{1}{315}\,\s^2 - \frac{4}{8505}\,\s^3\right)
        - \frac{8}{27\,\epsilon}\, L_\mu \nn \\ &- \frac{16}{27}\, L_\mu^2
        + \left(\frac{917}{486} - \frac{19\,\pi^2}{81} \right) + \left(
        \frac{172}{225} - \frac{2\,\pi^2}{27} \right)\s \nn \\ & + \left(
        -\frac{871057}{396900} + \frac{2\,\pi^2}{9} \right) \s^2 + \left(
        -\frac{83573783}{10716300} + \frac{64\,\pi^2}{81} \right) \s^3
        \Bigg],
\end{align}
\begin{align}
    F_{2,u}^{(7)}[b] = C_F \cdot \Bigg[& -\frac{5}{27\,\epsilon} -
        \frac{20}{27}\, L_\mu \nn \\ & + \frac{13}{162} +\left(
        \frac{25}{81} - \frac{\pi^2}{27} \right) \s + \left( \frac{118}{81} -
        \frac{4\,\pi^2}{27} \right) \s^2+ \left( \frac{10361}{2835} -
        \frac{10\,\pi^2}{27} \right)\s^3 \Bigg]. \nn 
\end{align}

%
%
\subsection{Diagrams \ref{fig:O1O2}d)}
The computation of the diagrams in Fig.~\ref{fig:O1O2}d) is by far the most
complicated piece in our entire calculation of the ${\cal O}(\alphas)$
corrections to the matrix element for $\bdll$. After various unsuccessful
attempts, we managed to obtain the result by using the dimension-shifting
method \cite{Tarasov} (see Appendix~\ref{section:tensor}), combined with the
method of partial integration (see Appendix~\ref{section:intbyparts}).  Since
we want to include the details of the actual calculation, we relegate them to
Appendix~\ref{section:diagram_d}. Here, we merely present the final results,
viz the contributions to the form factors, which read
\begin{align} 
  \label{eq:diagd}
  F_{2,u}^{(9)}[d] = C_F \cdot \Bigg[& \frac{2}{3\,\epsilon^2} +
  \frac{1}{\epsilon} \left( \frac{5}{3} - \frac{4\,L_s}{3} +
    \frac{8}{3}\,L_\mu + \frac{4\,i\pi}{3} \right) + \frac{16}{3}\,L_\mu^2
  + \frac{7}{6} -4\,L_s \\ & + \frac{2}{3}\,L_s^2 + 4\,i\pi -
  \frac{4\,i\pi}{3} \, L_s - \pi^2 + \left(\frac{20}{3} - \frac{16}{3}\,L_s
    + \frac{16\,i\pi}{3} \right) \, L_\m \nn
  \\ & + \left( \frac{2}{3} + \frac{2}{3}\,L_s - \frac{2}{3}\,L_s^2 -
    \frac{2\,i\pi}{3} + \frac{4\,i\pi}{3}\,L_s \right)\,\s + \left(
    \frac{2}{3} +2\,L_s -2\,i\pi\right) \, \s^2 \nn 
  \\ & + \left(\frac{2}{3} + \frac{10}{3}\, L_s + \frac{4}{3} \, L_s^2 -
    \frac{10\,i\pi}{3} - \frac{8\,i\pi}{3}\,L_s \right) \, \s^3 \Bigg], \nn \\
  \nn \\ F_{2,u}^{(7)}[d] = C_F \cdot \Bigg[& \frac{2}{3\,\e} + \frac{7}{3} +
  \frac{8}{3}\, L_\mu - \left(\frac{1}{3} - \frac{1}{3}\, L_s - \frac{1}{3}\,
    L_s^2 + \frac{i\pi}{3} + \frac{2\,i\pi}{3}\, L_s \right)\,\s
 \\ & -\left(\frac{1}{3} + \frac{1}{3}\,L_s - \frac{1}{3}\, L_s^2
  -\frac{i\pi}{3} + \frac{2\,i\pi}{3}\, L_s\right)\, \s^2 - \left(\frac{1}{3}
  + L_s - i\pi \right)\, \s^3 \Bigg] \nn.
\end{align}

\subsection{Diagrams \ref{fig:O1O2}c)}
The calculation of this diagram can be done in a very simple and efficient
way. We add the two subdiagrams and integrate out the loop momentum of the
virtual gluon. Next we integrate over the remaining loop momentum, being left
with a four dimensional Feynman parameter integral. After introducing a single
Mellin-Barnes representation of the occurring denominator, the parameter
integrals can all be performed. At this level, the result contains Euler
Beta-functions involving the Mellin-Barnes parameter. Finally, the
Mellin-Barnes integral can be resolved applying the residue theorem, which
naturally leads to an expansion in the parameter $\s$.  The contribution of
the diagrams in Fig. \ref{fig:O1O2}c) to the form factors reads
\begin{align}
    F_{2,u}^{(9)}[c] = C_F \cdot \Bigg[& \frac{2}{3\,\epsilon^2} +
        \frac{1}{\epsilon} \left( \frac{5}{3} - \frac{4\,L_s}{3} +
        \frac{8}{3}\,L_\mu + \frac{4\,i\pi}{3} \right) + \frac{16}{3}\,L_\mu^2
        \\ & + \frac{1}{2} - 6\,L_s + \frac{2}{3}\,L_s^2 + \frac{10\,i\pi}{3}
        - \frac{8\,i\pi}{3}\,L_s - \frac{5\,\pi^2}{3} \nn \\ & + \left(
        \frac{4}{3} - 4\,L_s + \frac{2}{3}\,L_s^2 + \frac{2\,\pi^2}{9} \right)
        \s + \left( -1-2\,L_s + \frac{2}{3}\,L_s^2 + \frac{2\,\pi^2}{9}
        \right) \s^2 \nn \\ & +\left( -\frac{41}{27} - \frac{10}{9}\,L_s
        + \frac{2}{3}\,L_s^2 + \frac{2\,\pi^2}{9} \right)\s^3 + \left(
        \frac{20}{3} + \frac{16\,i\pi}{3} - \frac{16}{3}\,L_s \right) L_\mu
        \Bigg], \nn \\ \nn \\ F_{2,u}^{(7)}[c] = C_F \cdot \Bigg[&
        \frac{1}{3\,\epsilon} + \frac{5}{2} + \frac{2\,i\pi}{3} + \left(
        \frac{2\,L_s}{3} - \frac{L_s^2}{3} - \frac{\pi^2}{9}\right) \s +
        \left( \frac{2}{3} - \frac{L_s^2}{3}-\frac{\pi^2}{9} \right) \s^2 \\ &
        + \left( \frac{5}{6} - \frac{L_s}{3} - \frac{L_s^2}{3} -
        \frac{\pi^2}{9} \right) \s^3 + \frac{4}{3}\,L_\mu \Bigg]. \nn
\end{align}
We also performed the calculation of this diagram in two different, more
complicated ways, namely by
\begin{itemize}
\item using the building block $J_{\a\b}$ given in \cite{AAGW} and then
  introducing a double Mellin-Barnes representation,
\item using the dimension-shifting and integration-by-parts techniques as
  explained when discussing the computation of the diagrams in
  Fig.~\ref{fig:O1O2}d) (see also Appendices~\ref{section:tensor} and
  \ref{section:intbyparts}).
\end{itemize}
We found that all
three calculations yield the same result and thus serve as an excellent check
for the dimension-shifting approach and for 
the very complicated double Mellin-Barnes calculation.

\subsection{Diagrams \ref{fig:O1O2}e)}
The diagrams in \Fref{fig:O1O2}e) may again be solved in two ways. The first
way is to use the large external momentum expansion technique
\cite{Smirnov}. The second possibility is to apply the dimension-shifting and
integration-by-parts procedure \cite{Tarasov} also for this diagram.  We do
without presenting the calculation and merely give the results for the
contributions to the form factors.
\begin{align}
    F_{2,u}^{(9)}[e] &= C_F \cdot \Bigg[ \, -\frac{2}{3}
    \left(\frac{1}{\epsilon} + 4\,L_\mu \right) - \frac{49}{9} -
    \frac{4\,i\pi}{3} + \frac{4}{3}\,L_s + \frac{16}{3} \, \zeta(3)
    \Bigg], \\ F_{2,u}^{(7)}[e] &= \, 0. \nn
\end{align}

%
%
\subsection{$\bm{\ordo{\alpha_s}}$ counterterms to
    $\bm{\bra d\,\ell^+\ell^-|O_{1,2}^{u,c}|b\ket}$}
 \label{countertermso2}
So far, we have calculated the two-loop matrix elements $\bra d\, \ell^+
\ell^- |C_i \, O_i^q|b \ket$ ($i=1,2$; $q=u,c$). As the operators mix under
renormalization, there are additional contributions proportional to
$C_i$. These counterterms arise from the matrix elements of the operators
\begin{equation}
    \sum_{j=1}^{2} \delta Z_{ij} (O_j^u+O_j^c) + \sum_{j=3}^{10} \delta Z_{ij}
    O_j + \sum_{j=11}^{12} \delta Z_{ij} (O_j^u+O_j^c), \quad i=1,2 ,
\end{equation}
where the operators $O_1$--$\,O_{10}$ are given in
Eq.~(\ref{oper}). $O_{11}^{u,c}$ and $O_{12}^{u,c}$ are evanescent
operators, i.e. operators which vanish in $d=4$ dimensions. In principle, there
is some freedom in the choice of the evanescent operators. However, as we want
to combine our matrix elements with the Wilson coefficients calculated by
Bobeth et al.~\cite{Bobeth:2000}, we have to use the same
definitions:
\begin{align}
    O_{11}^u & = \left( \bar d_L \gamma_\mu \gamma_\nu \gamma_\sigma T^a u_L
             \right) \left( \bar u_L \gamma^\mu \gamma^\nu \gamma^\sigma T^a
             b_L \right) - 16 \, O_1^u \, ,\nn \\ O_{12}^u & = \left(
             \bar d_L \gamma_\mu \gamma_\nu \gamma_\sigma u_L \right) \left(
             \bar u_L \gamma^\mu \gamma^\nu \gamma^\sigma b_L \right) - 16 \,
             O_2^u \, ,\\ O_{11}^c & = \left( \bar d_L \gamma_\mu \gamma_\nu
             \gamma_\sigma T^a c_L \right) \left( \bar c_L \gamma^\mu
             \gamma^\nu \gamma^\sigma T^a b_L \right) - 16 \, O_1^c \,
             ,\nn \\ O_{12}^c & = \left( \bar d_L \gamma_\mu \gamma_\nu
             \gamma_\sigma c_L \right) \left( \bar c_L \gamma^\mu \gamma^\nu
             \gamma^\sigma b_L \right) - 16 \, O_2^c \, . \nn
\end{align}
The operator renormalization constants $Z_{ij} = \delta_{ij} + \delta Z_{ij}$
are of the form
\begin{equation}
    \delta Z_{ij} = \frac{\alpha_s}{4\, \pi} \left( a_{ij}^{01} +
        \frac{1}{\epsilon}\, a_{ij}^{11}\right) + \frac{\alpha_s^2}{(4\,
        \pi)^2} \left( a_{ij}^{02} + \frac{1}{\epsilon}\, a_{ij}^{12} +
        \frac{1}{\epsilon^2}\, a_{ij}^{22}\right) + \ordo{\alpha_s^3}.
\end{equation}
The coefficients $a_{ij}^{lm}$ needed for our calculation we take from
Refs.~\cite{AAGW,Bobeth:2000} and list them for $i=1,2$ and
$j=1,...,12$:
\begin{align}
    \hat{a}^{11}= & \left(
    \begin{array}{cccccccccccc}
        -2& \frac{4}{3}&0& -\frac{1}{9} &0&0 & 0 &0& -
        \frac{16}{27}&0&\frac{5}{12}& \frac{2}{9} \\ \\ 6& 0&0 & \frac{2}{3}
        &0&0&0 &0& -\frac{4}{9}& 0& 1& 0
    \end{array}
    \right), \\ \nn \\ & \begin{array}{lll} a^{12}_{17} =
    -\frac{58}{243}\,, \hspace{0cm} & a^{12}_{19} = -\frac{64}{729}\,,
    \hspace{0cm} & a^{22}_{19} = \frac{1168}{243}\,, \\ \\ a^{12}_{27} =
    \frac{116}{81}\,, \hspace{0.cm} & a^{12}_{29} = \frac{776}{243}\,,
    \hspace{0.cm} & a^{22}_{29} = \frac{148}{81}\,.
    \end{array}
\end{align}
We denote the counterterm contributions to $\bdll$ which are due to the mixing
of $O_1^u$ or $O_2^u$ into four-quark operators by $F_{i,u \to \rm{4
quark}}^{\rm{ct}(7)}$ and $F_{i,u \to \rm{4 quark}}^{\rm{ct}(9)}$. They can be
extracted from the equation
\begin{equation}
    \label{fi4quark}
    \sum_{j} \left( \frac{\alpha_s}{4\,\pi}\right) \frac{1}{\epsilon} \,
    a_{ij}^{11} \bra d\, \ell^+ \ell^-|O_j^u|b \ket_{\text{1-loop}} = - \left(
    \frac{\alpha_s}{4\,\pi}\right) \left[ F_{i,u \to \rm{4
    quark}}^{\rm{ct}(7)} \bra \widetilde{O}_7\ket_{\text{tree}} + F_{i,u \to
    \rm{4 quark}}^{\rm{ct}(9)} \bra \widetilde{O}_9\ket_{\text{tree}} \right],
\end{equation}
where $j$ runs over the four-quark operators. The operators $O_j^u$ are
understood to be identified with $O_j$ for $j=3,4,5,6$.  As certain entries of
$\hat{a}^{11}$ are zero, only the one-loop matrix elements of $O_1^{u,c}$,
$O_2^{u,c}$, $O_4^{u,c}$, $O_{11}^{u,c}$ and $O_{12}^{u,c}$ are needed. In
order to keep the presentation transparent, we relegate their explicit form to
Appendix \ref{appendix:oneloop}. We do not repeat the renormalization of the
$O_1^c$ and $O_2^c$ contributions at this place and refer to \cite{AAGW}.

There is a counterterm related to the two-loop mixing of $O_i^u$ ($i=1,2$)
into $O_7$, followed by taking the tree-level matrix element $\bra d \,\ell^+
\ell^-|O_7|b\ket$.  Denoting the corresponding contribution to the counterterm
form factors by $F_{i,u \to 7}^{\rm{ct}(7)}$ and $F_{i,u \to 7}^{\rm{ct}(9)}$,
we obtain
\begin{equation}
    F_{i,u \to 7}^{\rm{ct}(7)} = - \frac{a_{i7}^{12}}{\epsilon} 
\, , \quad 
    F_{i,u \to 7}^{\rm{ct}(9)} = 0.
\end{equation}

The counterterms which are related to the mixing of $O_i^u$ ($i=1,2$) into
$O_9$ can be split into two classes: The first class consists of the one-loop
mixing $O_i^u \to O_9$, followed by taking the one-loop corrected matrix
element of $O_9$. It is obvious that this class contributes to the
renormalization of diagram \ref{fig:O1O2}f), which we take into account
when discussing the virtual corrections to $O_9$. We proceed in the same way
with the corresponding counterterm.

The second class of counterterm contributions due to $O_i^u \to O_9$ mixing is
generated by two-loop mixing of $O_2^u$ into $O_9$ as well as by one-loop
mixing and one-loop renormalization of the $g_s$ factor in the definition of
the operator $O_9$. We denote the corresponding contribution to the
counterterm form factors by $F_{i,u \to 9}^{\rm{ct}(7)}$ and $F_{i,u \to
9}^{\rm{ct}(9)}$. We obtain
\begin{equation}
    F_{i,u \to 9}^{\rm{ct}(9)} = -\left(\frac{a_{i9}^{22}}{\epsilon^2} +
    \frac{a_{i9}^{12}}{\epsilon} \right) - \frac{a_{i9}^{11} \,
    \beta_0}{\epsilon^2} \, , \quad F_{i,u \to 9}^{\rm{ct}(7)} = 0,
\end{equation}
where we used the renormalization constant $Z_{g_s}$ given by
\begin{equation}
\label{eq:Zg}
    Z_{g_s} = 1 - \frac{\alpha_s}{4\pi} \, \frac{\beta_0}{2} \,
    \frac{1}{\epsilon} \, , \quad \beta_0 = 11 - \frac{2}{3} N_f \,, \quad N_f
    = 5.
\end{equation}

The total counterterms $F_{i,u}^{\rm{ct}(j)}$ ($i=1,2$; $j=7,9$), which
renormalize diagrams \ref{fig:O1O2}a)--\ref{fig:O1O2}e), are given
by
\begin{equation}
    F_{i,u}^{\rm{ct}(j)} = F_{i,u \to \rm{4 quark}}^{\rm{ct}(j)} + 
F_{i,u \to 7}^{\rm{ct}(j)} +
F_{i,u \to 9}^{\rm{ct}(j)} \, .
\end{equation}
Explicitly they read
\begin{align}
    \nn F_{2,u}^{\rm{ct}(9)} = & -F_{2,u, \,\text{div}}^{(9)} -
    \frac{8}{25515} \left[2870 - 6300\,\pi^2 - 420\,i \pi + 126\,\s -
    \s^3\right] \\ \nn \\ & + \frac{8}{25515} \left[-420 - 21420\,i\pi +
    252\,\s + 27\,\s^2 + 4\,\s^3 \right] L_{\mu} \\ \nn \\ \nn &
    -\frac{136}{81}\, L_s^2 + \left[ \frac{16}{243} (-2+51\,i\pi) +
    \frac{544}{81}\,L_\mu\right] L_s -\frac{512}{81}\,L_\mu^2, \\ \nn \\
    \nn F_{2,u}^{\rm{ct}(7)} = & -F_{2,u, \,\text{div}}^{(7)} +
    \frac{2}{2835} \left(840\, L_{\mu} + 70\,\s + 7\,\s^2 + \s^3 \right),
\end{align}
\begin{align}
    F_{1,u}^{\rm{ct}(9)} = & -F_{1,u, \,\text{div}}^{(9)} + \frac{4}{76545}
        \left[ 59570 - 6300\,\pi^2 + 33600\,i\pi + 126\,\s - \s^3 \right]
        \nn \\ \nn \\ & + \frac{4}{76545} \left[68460 +
        21420\,i\pi -252\s -27 \s^2 -4\s^3\right]\,L_\mu
        \\ \nn \\ & + \frac{68}{243}\,L_s^2 - \left[
        \frac{8}{729} (160 + 51\,i\pi) + \frac{272}{243}\,L_\mu\right] L_s
        +\frac{256}{243}\,L_\mu^2, \nn \\ \nn \\ \nn
        F_{1,u}^{\rm{ct}(7)} = & -F_{1,u, \,\text{div}}^{(7)} - \frac{1}{8505}
        \left(840\, L_{\mu} + 70\, \s + 7\,\s^2 + \s^3 \right).
\end{align}
The quantities $F_{i,u, \,\text{div}}^{(j)}$ $(i=1,2;~j=7,9)$ compensate
the divergent parts of the form factors associated with the virtual
corrections to $O_{1,2}^u$. They are given
by
\begin{align}
    F_{2,u, \,\text{div}}^{(9)} & = \frac{128}{81\,\epsilon^2} +
        \frac{4}{25515\,\epsilon} \big[ 20790 + 21420\,i\pi - 252\,\s -
        27\,\s^2 - 4\,\s^3 \big] + \frac{16}{81\,\epsilon}(32\,L_\mu -
        17\,L_s), \nn \\ \nn \\ F_{2,u, \,\text{div}}^{(7)} & =
        \frac{92}{81\,\epsilon} \, , \nn \\ \\ F_{1,u,
        \,\text{div}}^{(9)} & = -\frac{64}{243\,\epsilon^2} -
        \frac{2}{76545\,\epsilon} \big[ 71820 + 21420\,i\pi - 252\,\s -
        27\,\s^2 - 4\,\s^3 \big] - \frac{8}{243\,\epsilon}(32\,L_\mu -
        17\,L_s), \nn \\ \nn \\ F_{1,u, \,\text{div}}^{(7)} & =
        -\frac{46}{243\,\epsilon} \, . \nn
\end{align}

As mentioned before, we will take diagram \ref{fig:O1O2}f) into account
only in \Sref{Section:O789}.  The same holds for the counterterms
associated with the $b$- and $d$-quark wave function renormalization and, as
stated earlier in this subsection, the $\ordo{\alpha_s}$ correction to the
matrix element of $\delta Z_{i9} O_9$. The sum of these contributions is
\[
   \delta \bar Z_\psi \bra O_i^u \ket_{\text{1-loop}} + \frac{\alpha_s}{4\pi}
   \,\frac{a_{i9}^{11}}{\epsilon} \, \left[ \delta \bar Z_\psi \bra O_9
   \ket_{\text{tree}} + \bra O_9 \ket_{\text{1-loop}} \right], \quad \delta
   \bar Z_\psi = \sqrt{ Z_\psi(m_b) Z_\psi(m_d)} - 1,
\]
and provides the counterterm that renormalizes diagram
\ref{fig:O1O2}f). We use on-shell renormalization for the external $b$-
and $d$-quark. In this scheme the field strength renormalization constants are
given by
\begin{equation}
\label{quarkren}
    Z_\psi(m) = 1 - \frac{\alpha_s}{4\pi} \, \frac{4}{3} \,
        \left(\frac{\mu}{m}\right)^{2\epsilon} \left( \frac{1}{\epsilon} +
        \frac{2}{\epsilon_\IR} + 4 \right).
\end{equation}
So far, we have discussed the counterterms which renormalize the
$\ordo{\alpha_s}$ corrected matrix elements $\bra d\, \ell^+ \ell^-|O_i^u|b
\ket$ ($i=1,2$). The corresponding one-loop matrix elements [of
$\ordo{\alpha_s^0}$] are renormalized by adding the counterterms
\[
    \frac{\alpha_s}{4\,\pi} \, \frac{a_{i9}^{11}}{\epsilon} \, \bra O_9
    \ket_\text{tree} \, .
\]

%
%
\subsection{Renormalized form factors of $\bm{O_1^u}$ and $\bm{O_2^u}$}
\label{section:renormo12u}
We now have all ingredients necessary to present the renormalized form factors
associated with the operators $O_1^u$ and $O_2^u$.  We stress again that only
the contributions of the diagrams~\ref{fig:O1O2}a)-e) and the counterterms
discussed in Subsection~\ref{countertermso2} are accounted for in the result
below. Diagram~\ref{fig:O1O2}f) and associated counterterms will be included
in the discussion of the virtual corrections to $O_9$. We decompose the
renormalized matrix elements of $O_i$ ($i=1,2$) as
\begin{equation}
    \bra d\,\ell^+\ell^-|O_i^u|b\ket = - \frac{\alpha_s}{4\,\pi} \, \left[
    F_{i,u}^{(9)} \bra \widetilde{O}_9 \ket_{\text{tree}} + F_{i,u}^{(7)} \bra
    \widetilde{O}_7 \ket_{\text{tree}} \right],
\end{equation}
where the operators $\widetilde{O}_7$ and $\widetilde{O}_9$ are defined
in Eq. (\ref{Otilde}). The renormalized form factors read:
\begin{align}
  \label{renormo12u}
  F_{1,u}^{(7)} = & -\frac{833}{729} - \frac{208}{243}\,L_\m -
  \frac{40\,i\pi}{243} + \left(-\frac{2}{729} - \frac{58}{243}\,L_s
  +\frac{2\,i\pi}{27} + \frac{4\,i\pi}{27}\,L_s + \frac{8\,\pi^2}{243}
  \right)\, \s \nn \\
  & + \left(-\frac{1453}{3645} + \frac{14}{243}\,L_s - \frac{2\,i\pi}{27} +
  \frac{4\,i\pi}{27}\,L_s + \frac{14\,\pi^2}{243} \right)\, \s^2 \\
  & + \left(-\frac{4712}{5103} + \frac{68}{243}\, L_s + \frac{2}{27}\, L_s^2 -
  \frac{2\, i\pi}{9} + \frac{26\,\pi^2}{243}\right)\, \s^3, \nn \\
  F_{1,u}^{(9)} = & -\frac{1736}{243} + \frac{224}{81}\, L_s -
  \frac{2864}{729}\, L_\m + \frac{272}{243}\,L_s\,L_\m -
  \frac{256}{243}\,L_\m^2 \nn \\ & - \frac{520\,i\pi}{243} +
  \frac{64\,i\pi}{243}\,L_s -\frac{272\,i\pi}{243}\,L_\m +
  \frac{200\,\pi^2}{729} + \frac{256}{27}\,\zeta(3) \nn \\ & +
  \left(-\frac{388}{675} + \frac{20}{27}\,L_s + \frac{16}{1215}\,L_\m +
  \frac{4\,i\pi}{27} - \frac{8\,i\pi}{27}\,L_s - \frac{8\,\pi^2}{243} \right)
  \, \s \\ &+
  \left(\frac{1018057}{1786050} + \frac{4}{243}\,L_s -\frac{4}{27}\,L_s^2 +
  \frac{4}{2835}\, L_\m + \frac{4\,i\pi}{9} - \frac{8\,\pi^2}{81} \right) \,
  \s^2 \nn \\ &
  + \left(\frac{92876363}{48223350} - \frac{344}{729}\,L_s -
  \frac{4}{9}\,L_s^2 + \frac{16}{76545}\,L_\m + \frac{20\,i\pi}{27} +
  \frac{16\,i\pi}{27}\, L_s - \frac{164\,\pi^2}{729} \right)\, \s^3, \nn \\
  F_{2,u}^{(7)} = & \frac{1666}{243} + \frac{416}{81}\,L_\m +
  \frac{80\,i\pi}{81} + \left(\frac{4}{243} + \frac{116}{81}\,L_s
  -\frac{4\,i\pi}{9} - \frac{8\,i\pi}{9}\,L_s - \frac{16\,\pi^2}{81} \right)\,
  \s \nn \\ & + \left(\frac{2906}{1215} - \frac{28}{81}\,L_s +
  \frac{4\,i\pi}{9} - \frac{8\,i\pi}{9}\,L_s - \frac{28\,\pi^2}{81} \right)\,
  \s^2 \\ & + \left(\frac{9424}{1701} - \frac{136}{81}\, L_s - \frac{4}{9}\,
  L_s^2 + \frac{4\,i\pi}{3} - \frac{52\,\pi^2}{81} \right)\, \s^3, \nn \\ \nn
  F_{2,u}^{(9)} = & -\frac{380}{81} - \frac{304}{27}\,L_s + \frac{3136}{243}\,
  L_\m - \frac{544}{81}\,L_s\,L_\m + \frac{512}{81}\,L_\m^2 \nn \\ & +
  \frac{608\,i\pi}{81} - \frac{128\,i\pi}{81}\,L_s +
  \frac{544\,i\pi}{81}\,L_\m - \frac{400\, \pi^2}{243} +
  \frac{64}{9}\,\zeta(3) \nn \\ & + \left(\frac{776}{225} - \frac{40}{9}\,L_s
  - \frac{32}{405}\,L_\m - \frac{8\,i\pi}{9} + \frac{16\,i\pi}{9}\,L_s +
  \frac{16\,\pi^2}{81} \right)\, \s \\ & + \left(-\frac{1018057}{297675} -
  \frac{8}{81}\,L_s + \frac{8}{9}\, L_s^2 - \frac{8}{945}\,L_\m -
  \frac{8\,i\pi}{3} + \frac{16\,\pi^2}{27} \right) \, \s^2 \nn \\ & +
  \left(-\frac{92876363}{8037225} + \frac{688}{243}\,L_s + \frac{8}{3}\,L_s^2
  - \frac{32}{25515}\,L_\m - \frac{40\,i\pi}{9} - \frac{32\,i\pi}{9}\,L_s +
  \frac{328\, \pi^2}{243} \right)\, \s^3, \nn
\end{align}
with
\[
    L_s = \ln(\s) \quad\text{and}\quad L_\mu =
    \ln\!\left(\!\frac{\mu}{m_b}\right) \, .
\]
As has been mentioned before, we only include terms up to $\ordo{\s^3}$ in the
result. We checked, however,  that the terms of order $\s^4$ are numerically
negligible. 
%
%
\section{Virtual Corrections to the Matrix Elements of the Operators
    $\bm{O_7}$, $\bm{O_8}$, $\bm{O_9}$ and $\bm{O_{10}}$}
\label{Section:O789}
The virtual corrections to the matrix elements of $O_7$, $O_8$, $O_9$ and
$O_{10}$ and their renormalization are discussed in detail in
Refs.~\cite{AAGW,letter}. For completeness we list the results of the
renormalized matrix elements.  They may all be decomposed according to
\[
    \bra d\,\ell^+\ell^- | C_i O_i | b \ket = \wtC_i^{(0)}
    \left(-\frac{\alpha_s}{4\,\pi}\right) \Big[ F_i^{(9)}
    \bra\wtO_9\ket_\text{tree} + F_i^{(7)} \bra\wtO_7\ket_\text{tree} \Big],
\]
where
\begin{align}
    \wtO_i = &\, \frac{\alpha_s}{4\,\pi}\,O_i,\, \nn \\[0.5ex]
    \wtC_7^{(0)} = &\, C_7^{(1)} ,\quad \wtC_8^{(0)} = C_8^{(1)} , \nn
    \\[0.5ex] \wtC_9^{(0)} = &\, \frac{4\,\pi}{\alpha_s} \left(C_9^{(0)} +
    \frac{\alpha_s}{4\,\pi} C_9^{(1)}\right) \quad\text{and}\quad
    \wtC_{10}^{(0)} = C_{10}^{(1)}. \nn
\end{align}

%
\subsection{Renormalized matrix element of $\bm{O_7}$}
The renormalized corrections to the form factors $F_7^{(9)}$ and $F_7^{(7)}$
are given by
\begin{eqnarray}
    \label{eq:FO7}
    F_7^{(9)} &=& -\frac{16}{3}\left(1 + \frac{1}{2}\, \s + \frac{1}{3}\, \s^2
    + \frac{1}{4}\, \s^3 \right), \\ \nn \\ F_7^{(7)} &=& \frac{32}{3}
    \, L_\mu + \frac{32}{3} + 8\, \s + 6\, \s^2 + \frac{128}{27}\, \s^3 +
    f_{\rm{inf}}\, .
\end{eqnarray}
The function $f_{\rm{inf}}$ collects the infrared- and collinear singular
parts:
\begin{equation}
    \label{eq:finf}
    f_{\rm{inf}} = \frac{\mubeps}{\epsilon_\IR} \, \frac{8}{3} \left(1 + \s +
    \frac{1}{2}\, \s^2 + \frac{1}{3}\, \s^3\right) +
    \frac{\mubeps}{\epsilon_\IR} \, \frac{4}{3} \, \ln(r) + \frac{2}{3} \ln(r)
    - \frac{2}{3} \ln^2(r),
\end{equation}
where $\epsilon_\IR$ and $r=\left(m_d^2/m_b^2\right)$ regularize the infrared-
and collinear singularities, respectively.

%
\subsection{Renormalized matrix element of the operator $\bm{O_8}$}
The renormalized corrections to the form factors of the matrix element of $O_8$
are
\begin{align}
    \label{eq:FO8}
    F_8^{(9)} & = \frac{104}{9} - \frac{32}{27}\, \pi^2 +
    \left(\frac{1184}{27} - \frac{40}{9}\, \pi^2\right) \s +
    \left(\frac{14212}{135} - \frac{32}{3}\, \pi^2 \right)\s^2 \\ \nn \\
    & + \left(\frac{193444}{945} - \frac{560}{27}\, \pi^2 \right) \s^3 +
    \frac{16}{9}\, L_s \left( 1 + \s + \s^2 + \s^3 \right), \nn
\end{align}
\begin{align}
    F_8^{(7)} & = - \frac {32}{9}\, L_\mu + \frac{8}{27}\, \pi^2 -
    \frac{44}{9} - \frac{8}{9}\, i \pi + \left(\frac{4}{3}\, \pi^2 -
    \frac{40}{3} \right) \s + \left(\frac{32}{9}\, \pi^2 -
    \frac{316}{9}\right)\s^2 \\ \nn \\ & + \left(\frac{200}{27}\, \pi^2
    - \frac{658}{9} \right) \s^3 - \frac {8}{9}\, L_s \left( \s + \s^2 + \s^3
    \right). \nn
\end{align}

%
\subsection{Renormalized matrix element of $\bm{O_9}$ and $\bm{O_{10}}$}
The renormalized matrix elements of $O_9$ and $O_{10}$, finally, are described
by the form factors
\begin{align}
    \label{eq:FO9}
    F_9^{(9)} & = \frac{16}{3} + \frac{20}{3}\, \s + \frac{16}{3}\, \s^2 +
    \frac{116}{27}\, \s^3 + f_{\rm{inf}}\, , \\ F_9^{(7)} & = -\frac{2}{3}\s
    \left(1 + \frac{1}{2}\, \s + \frac{1}{3}\, \s^2 \right), \\ F_{10}^{(9)} &
    = F_9^{(9)}, \vspace*{0.6cm} \\ \vspace*{0.3cm} F_{10}^{(7)} & =
    F_9^{(7)},
\end{align}
where $f_{\rm{inf}}$ is defined in Eq.~(\ref{eq:finf}).

The contribution of the renormalized diagrams~\ref{fig:O1O2}f), which have been
omitted so far, is properly included by modifying $\wtC_9^{(0)}$ as follows:
\[
    \wtC_9^{(0)} \to \wtC_9^{(0,\text{mod})} = \wtC_9^{(0)} -
    \frac{1}{\xi_t} \left(C_2^{(0)} + \frac{4}{3}\, C_1^{(0)}\right) \big(
    \xi_u H_0(0) + \xi_c H_0(z) \big).
\]
For $\s < 4\,z$ $(z=m_c^2/m_b^2)$ the loop function $H_0(z)$ can be expanded
in terms of $\s/(4\,z)$. We give the expansion of $H_0(z)$ as well as the
result for $H_0(0)$:
\begin{align}
    H_0(z) & = \frac{1}{2835} \left[ -1260 + 2520 \ln
    \!\left(\frac{\mu}{m_c}\right) + 1008 \left(\frac{\s}{4z}\right) +
    432 \left(\frac{\s}{4z}\right)^2 + 256
    \left(\frac{\s}{4z}\right)^3 \right], \nn \\ \\ H_0(0) & =
    \frac{8}{27} -\frac{4}{9}\,\ln(\s) + \frac{4\,i\pi}{9} + \frac{8}{9}\,L_\m\nn.
\end{align}
%
%
\section{Corrections to the Decay Width $\bm{B \rightarrow X_d \ell^+\ell^-}$}
\label{section:decay_width}
The decay width differential in $\s$ can be written as
\begin{eqnarray}
\label{rarewidth}
    \frac{d\Gamma(b\to X_d \ell^+\ell^-)}{d\s} &=&
    \left(\frac{\alpha_{\text{em}}}{4\,\pi}\right)^2 \frac{G_F^2\,
    m_{b,\pole}^5\,|\xi_t|^2} {48\,\pi^3}(1-\s)^2 \Bigg\{\left(1 + 2\,\s
    \right) \left (\left |\widetilde C_9^{\eff}\right|^2 + \left |\widetilde
    C_{10}^{\eff}\right |^2 \right ) \nn \\ && + 4(1+2/\s)\left|
    \widetilde C_7^{\eff}\right |^2 + 12 \, \mbox{Re}\left (\widetilde
    C_7^{\eff} \widetilde C_9^{\eff*}\right ) \Bigg\} \nn \\&&
    +\frac{d\Gamma^{\rm Brems,\,A}}{d\s}+\frac{d\Gamma^{\rm Brems,\,B}}{d\s}.
\end{eqnarray}
The last two terms in \Eref{rarewidth} correspond to certain finite
bremsstrahlung contributions specified in Appendix \ref{appendix:brems}. Their
result can also be found in this appendix.  All other corrections have
been absorbed into the effective Wilson coefficients $\widetilde{C}_7^{\eff}$,
$\widetilde{C}_9^{\eff}$ and $\widetilde{C}_{10}^{\eff}$. We follow
\cite{AAGW,letter,Bobeth:2000} and write the effective Wilson coefficients as
\begin{align}
  \label{effcoeff}
    \nn \widetilde C_9^{\eff} = & \left( 1 + \frac{\alpha_s(\mu)}{\pi}
    \omega_9 (\s)\right ) \big( A_9 -\frac{\xi_c}{\xi_t}\, T_{9a} \,
    h(z,\s) - \frac{\xi_u}{\xi_t}\, T_{9a} \, h(0,\s) +T_{9b} \,
    h(z,\s) \\ \nn &  + U_9 \, h(1,\s) +
    W_9 \, h(0,\s) \big) \\ \nn & + \frac{\alpha_{s}(\mu)}{4\pi}\left(
    \frac{\xi_u}{\xi_t} \left(C_1^{(0)} F_{1,u}^{(9)} + C_2^{(0)}
    F_{2,u}^{(9)}\right) + \frac{\xi_c}{\xi_t} \left(C_1^{(0)}
    F_{1,c}^{(9)} + C_2^{(0)} F_{2,c}^{(9)}\right) - A_8^{(0)}
    F_8^{(9)}\right),\nn \\ \nn \\
    \widetilde C_7^{\eff} = & \left
    (1+\frac{\alpha_s(\mu)}{\pi} \omega_7 (\s)\right )A_7 \\ \nn & +
    \frac{\alpha_s(\mu)}{4\pi}\left( \frac{\xi_u}{\xi_t}
    \left(C_1^{(0)} F_{1,u}^{(7)} + C_2^{(0)} F_{2,u}^{(7)}\right) +
    \frac{\xi_c}{\xi_t} \left(C_1^{(0)} F_{1,c}^{(7)} + C_2^{(0)}
    F_{2,c}^{(7)}\right) - A_8^{(0)} F_8^{(7)}\right), \nn \\ \nn
    \\ \nn \widetilde C_{10}^{\eff} = & \left( 1 +
    \frac{\alpha_s(\mu)}{\pi} \omega_9(\s) \right) A_{10} \nn ,
\end{align}
where we have provided the necessary modification to account for the CKM
structure of $\bdll$. The renormalized form factors $F_{1,u}^{(7,9)}$ and
$F_{2,u}^{(7,9)}$ and can be found in \Sref{section:renormo12u} while the
renormalized form factors $F_{1,c}^{(7,9)}$, $F_{2,c}^{(7,9)}$ and
$F_{8}^{(7,9)}$ are given in \cite{AAGW,letter}.  The functions $\omega_7(\s)$
and $\omega_9(\s)$ encapsulate the interference between the tree-level and the
one-loop matrix elements of $O_7$ and $O_{9,10}$ and the corresponding
bremsstrahlung corrections, which cancel the infrared- and collinear
divergences appearing in the virtual corrections. When calculating the decay
width (\ref{rarewidth}), we retain only terms linear in $\alpha_s$ (and thus
in $\omega_7$, $\omega_9$) in the expressions for $|\widetilde C_7^{\eff}|^2$,
$|\widetilde C_9^{\eff}|^2$ and $|\widetilde C_{10}^{\eff}|^2$.  Accordingly,
we drop terms of $\ordo{\alpha_s^2}$ in the interference term $\Re \left
(\widetilde C_7^{\eff} \widetilde C_9^{\eff*}\right )$, too, where by
construction one has to make the replacements $\omega_9 \to \omega_{79}$ and
$\omega_7 \to \omega_{79}$ in this term. The function $\omega_9$ has already
been calculated in \cite{Bobeth:2000}, where also the exact expression for
$h(\s,z)$ can be found. For the functions $\omega_7$ and $\omega_{79}$ and
more information on the cancellation of infrared- and collinear divergences we
refer to \cite{AAGW}.

The auxiliary quantities $A_7$, $A_9$, $A_{10}$, $T_{9a}$, $T_{9b}$, $U_9$ and
$W_9$ are the following linear combinations of the Wilson coefficients
$C_i(\mu)$:
\begin{align}
    \label{ATUW}
    A_7 =&\, \frac{4\, \pi}{\alpha_s(\mu)} \, C_7(\mu) - \frac{1}{3} \,
    C_3(\mu) - \frac{4}{9} \, C_4(\mu) - \frac{20}{3} \, C_5(\mu) -
    \frac{80}{9} \, C_6(\mu), \nn \\ A_8 =&\, \frac{4\,
      \pi}{\alpha_s(\mu)} \, C_8(\mu) + C_3(\mu) - \frac{1}{6} \,
    C_4(\mu) + 20 \, C_5(\mu) - \frac{10}{3} \, C_6(\mu), \nn \\
    A_9 =&\, \frac{4 \pi}{\alpha_s(\mu)} \, C_9(\mu) + \frac{4}{3} \,
    C_3(\mu) + \frac{64}{9} \, C_5(\mu) + \frac{64}{27} \,
    C_6(\mu)\nn \\ & + \left[ \frac{\xi_u +
        \xi_c}{-\xi_t} \left( C_1(\mu) \, \gamma_{19}^{(0)} +
        C_2(\mu) \, \gamma_{29}^{(0)} \right) + \sum_{i=3}^{6} \, C_i(\mu)
      \, \gamma_{i9}^{(0)} \right] \, \ln \! \left( \frac{m_b}{\mu}
    \right), \nn \\ A_{10} = &\, \frac{4\, \pi}{\alpha_s(\mu)}
    \, C_{10}(\mu), \\ T_{9a} =&\, \frac{4}{3} \, C_1(\mu) + C_2(\mu)\,,
    \nn \\ T_{9b} =&\, 6 \, C_3(\mu) + 60 \, C_5(\mu), \nn \\ U_9 =& -
    \frac{7}{2} \, C_3(\mu) - \frac{2}{3} \,C_4(\mu) -38 \,C_5(\mu) -
    \frac{32}{3} \,C_6(\mu), \nn \\ W_9 =& - \frac{1}{2} \,
    C_3(\mu) - \frac{2}{3} \,C_4(\mu) -8 \,C_5(\mu) - \frac{32}{3}
    \,C_6(\mu).\nn
\end{align}
In these definitions we also include some diagrams induced by $O_{3,4,5,6}$
insertions, viz the $\ordo{\alpha_s^0}$ contributions, the diagrams of
topology \ref{fig:O1O2}f) and those bremsstrahlung diagrams where the
gluon is emitted from the $b$- or $d$-quark line (cf \cite{Asa2}).

For completeness, we give in Table~\ref{tableA} numerical values 
for $C_1$, $C_2$, $A_7$, $A_8$, $A_9$, $A_{10}$, $T_{9a}$, $T_{9b}$,
$U_9$ and $W_9$ at three different values of the renormalization scale
$\mu$. We note that the recently calculated contributions \cite{Gambino} 
to the anomalous dimension matrix which correspond to the three-loop 
mixings of the four-quark operators into $O_9$ have been included
by adopting the procedure described in Appendix~\ref{appendix:wilson}.
As can be seen in Table~\ref{tableA}, 
some of the entries have a very small amount of
significant digits. In our numerical analysis presented in
\Sref{section:phenom} we work with a much higher accuracy.
\begin{table}[htb]
    \begin{center}
      \renewcommand{\st}{\rule[-2.5mm]{0mm}{8mm}}
      \def\lb{\raisebox{0.5mm}{\big(}} \def\rb{\raisebox{0.5mm}{\big)}}
    \begin{tabular*}{\textwidth}{c@{\hspace*{1.5cm}}c@{\hspace*{1.5cm}}c@{\hspace*{1.5cm}}c}
        \hline\hline \st & $ \mu=2.5$ GeV & $ \mu=5$ GeV & $ \mu=10$ GeV \\
        \hline \st$\alpha_s $ & $ 0.267 $ & $ 0.215 $ & $ 0.180 $ \\ \st$\lb
        C_1^{(0)},~C_1^{(1)}\rb $ & $ (-0.696,~0.240) $ & $ (-0.486,~0.206) $
        & $(-0.326,~0.184) $ \\ \st$\lb C_2^{(0)},~C_2^{(1)}\rb $ & $
        (1.046,~-0.276) $ & $ (1.023,~-0.017) $ & $ (1.011,~-0.010) $ \\
        \st$\lb A_7^{(0)},~A_7^{(1)}\rb $ & $ (-0.360,~0.032) $ &
        $(-0.321,~0.018) $ & $ (-0.287,~0.009) $ \\ \st$\lb
        A_8^{(0)},~A_8^{(1)}\rb $ & $ (-0.169,~-0.015) $ & $ (-0.153,~-0.013)
        $ & $ (-0.140,-0.012) $ \\ \st$\lb A_9^{(0)},~A_9^{(1)}\rb $ & $
        (4.241,~-0.091) $ & $(4.128,~0.066) $ & $ (4.131,~0.190) $ \\ \st$\lb
        T_{9a}^{(0)},~T_{9a}^{(1)}\rb $ & $ (0.118,~0.292) $ & $(0.376,~0.258)
        $ & $ (0.577,~0.235) $ \\ \st$\lb T_{9b}^{(0)},~T_{9b}^{(1)}\rb $ & $
        (-0.003,~-0.013) $ & $(-0.001,~-0.007) $ & $ (0.000,~-0.004) $\\
        \st$\lb U_9^{(0)},~U_9^{(1)}\rb $ & $ (0.045,~0.023) $ &
        $(0.033,~0.015) $ & $ (0.022,~0.010) $ \\ \st$\lb
        W_{9}^{(0)},~W_{9}^{(1)}\rb $ & $ (0.044,~0.016) $ & $(0.032,~0.012) $
        & $ (0.022,~0.009) $ \\ \st$\lb A_{10}^{(0)},~A_{10}^{(1)}\rb $ & $
        (-4.373,~0.135) $ & $(-4.373,~0.135) $ & $ (-4.373,~0.135) $ \\
        \hline\hline
    \end{tabular*}
    \end{center}
    \caption{Coefficients appearing in Eq.~(\ref{ATUW}) for $\mu =
      2.5$~GeV, $\mu = 5$~GeV and $\mu = 10$~GeV. For $\alpha_s(\mu)$ (in the
    $\overline{\mbox{MS}}$ scheme) we used the two-loop expression with 5
    flavors and $\a_s(m_Z) = 0.119$. The entries correspond to the pole top
    quark mass $m_t = 174 $~GeV. The matching for the top and for the charm
    contribution was performed at a scale of $120$~GeV and $80$~GeV,
    respectively \cite{Bobeth:2000}. The superscript $(0)$ refers to lowest
    order quantities, while the superscript $(1)$ denotes the correction terms
    of order $\alpha_s$, i.e. $X=X^{(0)}+X^{(1)}$ with
    $X=C,\,A,\,T,\,U,\,W$. Note that the contributions calculated recently in
    Ref.~\cite{Gambino} are included. These contributions only affect the
    entries for $A_9^{(1)}$.}
    \label{tableA}
\end{table}

\section{Phenomenological analysis}
\label{section:phenom}
As the main point of this paper is the {\it calculation} of the NNLL 
corrections to the process $b \to X_d \ell^+ \ell^-$, 
we keep the phenomenological analysis rather short. In the following
we investigate the impact of the NNLL corrections on three observables:
the branching ratio, the CP asymmetry and the normalized
forward-backward asymmetry. As our main point is to illustrate the differences
between NLL and NNLL results, we do not include power corrections (and/or
effects from resonances), postponing this to future studies.

Since the decay width given in \Eref{rarewidth} suffers from a large
uncertainty due to the factor $m_{b,\text{pole}}^5$, we follow common practice
and introduce the ratio
\begin{eqnarray}
  \label{eq:rquark}
  R_{\text{quark}}(\s) & = & \frac{1}{\Gamma(b \to X_c \, e \,
  \bar{\nu}_e)} \frac{\frac{d\Gamma(b \to X_d \ell^+\,\ell^-)}
  {d\s}+\frac{d\Gamma(\bar{b} \to
  X_{\bar{d}}\ell^+\,\ell^-)}{d\s}}{2},
\end{eqnarray}
in which the factor $m_{b,\text{pole}}^5$ drops out. Note that we define
$R_{\text{quark}}(\s)$ as a charge-conjugate average as this is likely to be
the first quantity measured.  The expression for the semileptonic decay width
$\Gamma(b \to X_c \, e \, \bar{\nu}_e)$ is as follows:
\begin{eqnarray}
  \label{eq:semileptonic}
  \Gamma(b \to X_c \, e \, \bar{\nu}_e) &=& \frac{G_F^2\,m_{b,\text{pole}}^5}
  {192\pi^3} |V_{cb}|^2 \cdot g\left(\frac{m_{c,\text{pole}}^2}
  {m_{b,\text{pole}}^2} \right) \cdot K\left(\frac{m_c^2}{m_b^2}\right),
\end{eqnarray}
where $g(z)\,=\, 1\,-8\,z\,+8\,z^3\,-\,z^4\,-\,12\,z^2\,\ln(z)$ is the phase
space factor, and
\begin{eqnarray}
  \label{eq:SM_qcd}
  K(z) &=& 1 - \frac{2\alpha_s(m_b)}{3\pi} \frac{f(z)}{g(z)}
\end{eqnarray}
incorporates the next-to-leading QCD correction to the semileptonic decay. The
function $f(z)$ has been given analytically in Ref.~\cite{Nir}: 
\begin{eqnarray}
\label{eq:ffun}
&&f(z) = -(1-z^2) \, \left( \frac{25}{4} - \frac{239}{3} \, z +
\frac{25}{4} \, z^2 \right) + z \, \ln(z) \left( 20 + 90 \, z
-\frac{4}{3} \, z^2 + \frac{17}{3} \, z^3 \right)
\nonumber \\
&& + z^2 \, \ln^2(z) \, (36+z^2)
+ (1-z^2) \, \left(\frac{17}{3} -\frac{64}{3} \, z + \frac{17}{3} \, z^2
\right) \, \ln (1-z) \nonumber \\
&& -4 \, (1+30 \, z^2 + z^4) \, \ln(z) \ln(1-z)
-(1+16 \, z^2 +z^4)  \left( 6 \, \mbox{Li}(z) - \pi^2 \right)
\nonumber \\
&& -32 \, z^{3/2} (1+z) \left[\pi^2 - 4 \, \mbox{Li}(\sqrt{z})+
4 \, \mbox{Li}(-\sqrt{z}) - 2 \ln(z) \, \ln \left(
\frac{1-\sqrt{z}}{1+\sqrt{z}} \right) \right] \, .
\end{eqnarray}
In the following analysis
we write the CKM parameters appearing in $b \to X_d \ell^+ \ell^-$ as 
(neglecting terms of ${\cal O}(\lambda^7)$)
\[ \xi_u = A \, \lambda^3 \, (\bar{\rho}- i \bar{\eta}), \quad
   \xi_t = A \, \lambda^3 \, (1 - \bar{\rho}+ i \bar{\eta}), \quad
   \xi_c = -\xi_u - \xi_t , 
\]
with $\bar{\rho} = \rho (1 - \lambda^2/2)$ and $\bar{\eta} = \eta (1 -
     \lambda^2/2)$ \cite{Ostermaier}.  For $V_{cb}$, appearing in the
     semileptonic decay width, we use $V_{cb}= A \lambda^2$.  Numerically, we
     set $A=0.81$, $\lambda=0.22$, $\bar{\rho}=0.22$ and
     $\bar{\eta}=0.35$. For the other input parameters we use
     $\alpha_s(m_Z)=0.119$, $m_t^{\rm{pole}}=174$~GeV,
     $\alpha_{\rm{em}}=1/133$, $m_b=4.8$~GeV, $m_c/m_b=0.29$, $m_W=80.41$~GeV,
     $m_Z=91.19$~GeV, and $\sin^2(\theta_W)=0.231$.

\begin{figure}[H]
\begin{center}
    \includegraphics[width=7.5cm]{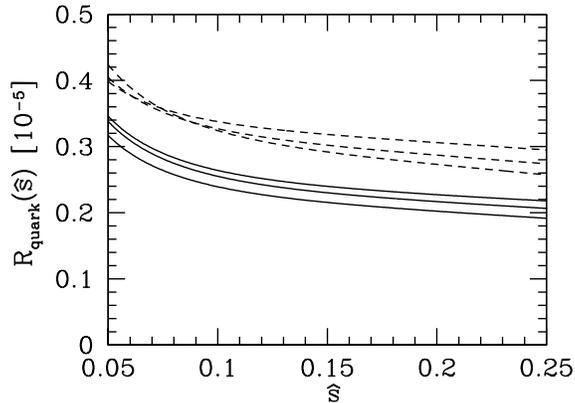}
    \vspace{0.5cm}
    \caption[]{$R_{\rm quark}(\s)$ as defined in Eq. (\ref{eq:rquark}).
    The solid lines show the NNLL
    result for $\mu=2.5,\,5.0,\,10.0$~GeV, whereas the dashed
    lines show the corresponding result in the NLL approximation.
    At $\s=0.25$ the highest (lowest) curve correspond to $\mu=10$~GeV
    ($\mu=$2.5~GeV) both in the NLL and NNLL case.}
\label{fig:Rquark}
\end{center}
\end{figure}

In \Fref{fig:Rquark} we show the $\mu$-dependence of $R_{\rm quark}(\s)$ for
$0.05 \leq \s \leq 0.25$.  The solid lines correspond to the NNLL results,
whereas the dashed lines represent the NLL results. We see that, going from
NLL to NNLL precision, $R_{\rm quark}(\s)$ is decreased throughout the entire
region by about $20-30\%$. Although the absolute uncertainty due to the
$\mu$-dependence decreases as well, the relative error remains roughly the
same.
As mentioned already in the introduction, 
the region $0.05 \le \s \le 0.25$ is free of resonances, as it lies below the
$J/\Psi$ threshold and above the $\rho$ and $\omega$ resonances.  
The contribution of this region 
to the decay width (normalized by $\Gamma(b \to X_c e \bar{\nu}_e)$)
is therefore well approximated by 
integrating $R_{\rm quark}(\s)$ over this interval. At NNLL precision, we
get
\begin{equation}
  \label{eq:int_rquark}
  R_{\rm quark} \,=\, \int\limits_{0.05}^{0.25}\!d\s\, R_{\rm quark}(\s) \,=\,
  \left(4.75 \pm 0.25 \right) \times 10^{-7}.
\end{equation}
The error is obtained by varying the scale $\mu$ between $2.5$~GeV and
$10$~GeV. The corresponding result in NLL precision is $R_{\rm quark} \,=\,
\left(6.29 \pm 0.21 \right) \times 10^{-7}$.  The renormalization scale
dependence therefore increases from $\sim\pm 3.4\%$ to $\sim\pm 5.3\%$.  The
reason for this increace can be understood from Fig.~\ref{fig:Rquark}: While
for $0.13 < \s < 0.25$ the $\mu$ dependence of $ R_{\rm quark}(\s)$ at NNLL
and NLL precision is similar, the $\mu$ dependence almost cancels in the NLL
case when integrating $\s$ between 0.05 and 0.13 due to the crossing of the
dashed lines in this interval. This cancellation does not happen in the NNLL
case, leading to a slightly larger $\mu-$dependence of $ R_{\rm quark}$ at
NNLL.
\begin{figure}[H]
\begin{center}
    \includegraphics[width=7.5cm]{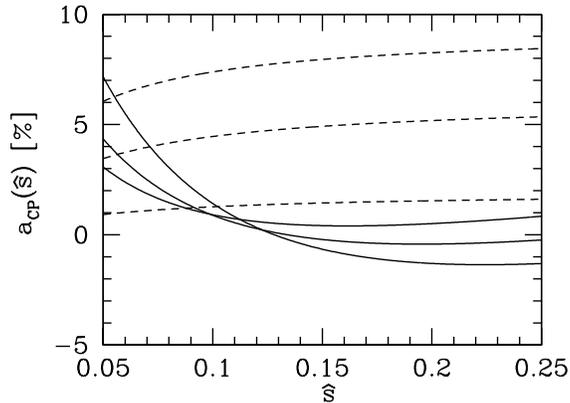}
    \vspace{0.5cm}
    \caption[]{CP asymmetry: The solid lines show the NNLL result for
    $\mu=2.5,\,5.0,\,10.0$~GeV, whereas the dashed lines show the
    corresponding result in the NLL approximation.
    At $\s=0.25$ the highest (lowest) curve correspond to $\mu=10$~GeV
    ($\mu=$2.5~GeV) both in the NLL and NNLL case.}
\label{fig:CP}
\end{center}
\end{figure}
As pointed out already, in the process $b \to X_d \ell^+ \ell^-$ the
contribution of the $u$-quark running in the fermion loop is, in contrast to
$b \to X_s \ell^+ \ell^-$, not Cabibbo-suppressed.  As a consequence, CP
violation effects are much larger in $b \to X_d \ell^+ \ell^-$.  The CP
asymmetry $a_{\rm CP}(\s)$ is defined as
\begin{eqnarray}
  \label{eq:asym}
  a_{\rm CP}(\s) & = & \frac{\frac{d\Gamma(b \to
  X_d \ell^+ \ell^-)}{d\s}-\frac{d\Gamma(\bar{b} \to
  X_{\bar{d}}  \ell^+\ell^-)}{d\s}}{\frac{d\Gamma(b \to
  X_d \ell^+ \ell^-)} {d\s}+\frac{d\Gamma(\bar{b} \to
  X_{\bar{d}} \ell^+\ell^-)}{d\s}} \, .
\end{eqnarray}
In \Fref{fig:CP} we show $a_{\rm CP}(\s)$ for $0.05\le\s\le0.25$. The solid
and dashed lines correspond to the NNLL and NLL results, respectively.  We
find several differences between the two results: The solid lines are much
closer together. Also they cross each other at $\s\approx 0.11$. Furthermore,
the NLL result clearly shows a positive CP asymmetry throughout the entire
$\s$ region considered, while the NNLL lines indicate that $a_{\rm CP}(\s)$
can be both positive and negative, depending on the value of $\s$.  Because of
that, it does not make much sense to quantify the relative error due to the
$\m$-dependence. The plot, however, clearly shows that the absolute
uncertainty is much smaller in the NNLL result. For NLL results, see also
Ref.~\cite{Ali:1998sf}.\\
We also give the averaged CP asymmetry $a_{\rm CP}$ in the region $0.05 \le
\s \le 0.25$, defined as
\begin{eqnarray}
  \label{eq:averaged_cp}
  a_{\rm CP} & = & \frac{\int\limits_{0.05}^{0.25}\!d\s\,\left(
  \frac{d\Gamma(b \to X_d \ell^+\ell^-)} {d\s}-\frac{d\Gamma(\bar{b} \to
  X_{\bar{d}} \ell^+\ell^-)}{d\s}\right)}{\int\limits_{0.05}^{0.25}\!d\s\,
  \left( \frac{d\Gamma(b \to X_d \ell^+\ell^-)} {d\s}+\frac{d\Gamma(\bar{b}
  \to X_{\bar{d}}\ell^+\ell^-)}{d\s}\right)} \, .
\end{eqnarray}
Varying $\mu$ between 2.5~GeV and 10~GeV one obtains the ranges
\begin{equation*}
  1.4 \% \le a_{\rm CP}^{\rm NLL} \le 7.7 \%, \quad ; \quad
  0.56 \% \le a_{\rm CP}^{\rm NNLL} \le  0.93 \% \, .
\end{equation*}

We now turn to the forward-backward asymmetry.  As for $R_{\rm quark}(\s)$,
we introduce a CP-averaged version of the normalized forward-backward
asymmetry, defined as
\begin{eqnarray}
  \label{eq:fbasym}
  \bar{A}_{\rm FB}(\s) &=&
  \frac{\int\limits_{-1}^{1}\,d\left(\cos\theta\right)\,{\rm sgn}(\cos\theta)
  \left(\frac{d^2\Gamma (b\to X_d
  \ell^+\ell^-)}{d\s\,d\left(\cos\theta\right)} + \frac{d^2\Gamma (\bar{b}\to
  X_{\bar{d}}
  \ell^+\ell^-)}{d\s\,d\left(\cos\theta\right)}\right)}{\frac{d\Gamma(b\to X_d
  \ell^+\ell^-)}{d\s}+\frac{d\Gamma (\bar{b}\to X_{\bar{d}}
  \ell^+\ell^-)}{d\s}} \, ,
\end{eqnarray}
where $\theta$ is the angle between the three-momenta of the positively
charged lepton $\ell^+$ and the $b$-quark in the rest frame of the lepton
pair.  The result of the integrals in the numerator of \Eref{eq:fbasym} for
the case $b \to X_s \ell^+ \ell^-$ can be found in \cite{Asatrian:2002va}. The
corresponding result for $b \to X_d \ell^+ \ell^-$ is, up to different
CKM-structures, the same.
\begin{figure}[H]
\begin{center}
    \includegraphics[width=7.5cm]{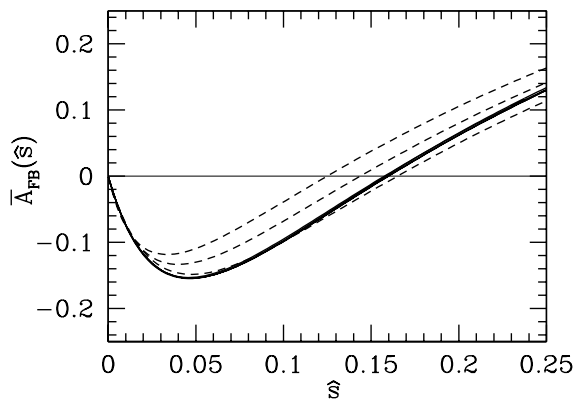}
    \vspace{0.5cm}
    \caption[]{CP-averaged normalized forward-backward asymmetry. 
    The solid lines show
    the NNLL result for $\mu=2.5,\,5.0,\,10.0$~GeV, whereas the
    dashed lines represent the corresponding result in the NLL approximation.}
 \label{fig:fbasym}
\end{center}
\end{figure}
In Fig. \ref{fig:fbasym} we illustrate the $\mu$-dependence of $\bar{A}_{\rm
FB}(\s)$ in the region $0 \le \s \le 0.25$. Again, the solid and dashed lines
represent the NNLL and the NLL results, respectively.  The reduction of the
$\m$-dependence going from NLL to NNLL precision is striking: one can clearly
distinguish the three dashed lines, whereas the NNLL lines are on top of each
other throughout the region. The position $\s_0$ at which the forward-backward
asymmetries vanish, is essentially free of uncertainties due to the variation
of $\m$: we find $\s_0^{\rm NNLL} = 0.158 \pm 0.001$. To NLL precision we get
$\s_0^{\rm NLL} = 0.145 \pm 0.020$.

\begin{figure}[H]
\begin{center}
    \includegraphics[width=7.5cm]{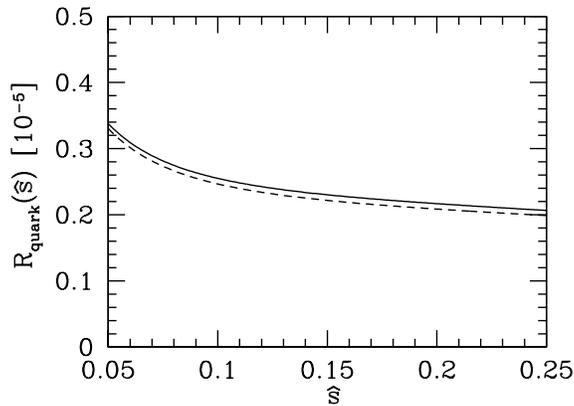}
    \vspace{0.5cm}
    \caption[]{$R_{\rm quark}(\s)$ for $\mu=5$~GeV. The solid line
    corresponds to matching top and charm contributions at $120$~GeV and 
    $80$~GeV, respectively. The dashed curve is obtained by matching both
    contributions at a scale of $80$~GeV.}
 \label{fig:matching}
\end{center}
\end{figure}

As a last illustration, we show in \Fref{fig:matching} the dependence of
$R_{\rm quark}(\s)$ on the matching scales. In all the previous plots we used
a matching scale of $120$~GeV for the top contribution and a matching scale of
$80$~GeV for the charm contribution. In \Fref{fig:matching} the solid
line corresponds to this scheme, while the dashed line is obtained by
matching both contributions at a scale of $80$~GeV. The difference between
the two schemes is between 2\% and 4\%.

\section{Summary}
\label{section:summary}
In this paper we presented the calculation of virtual and bremsstrahlung
corrections of $\ordo{\alphas}$ to the inclusive semileptonic decay $b \to X_d
\ell^+ \ell^-$.  Genuinely new calculations were necessary to attain the
virtual contributions of the operators $O_1^u$ and $O_2^u$. Some of the
diagrams (in particular diagrams \ref{fig:O1O2}d) turned out to be more
involved than the corresponding diagrams for the $c$-quark contributions. We
used dimension-shifting and integration-by-parts techniques to calculate
them. The main result of this paper, namely the $u$-quark contributions to the
renormalized form factors $F_{1,u}^{(7)}$, $F_{1,u}^{(9)}$, $F_{2,u}^{(7)}$,
and $F_{2,u}^{(9)}$, is given in \Sref{section:renormo12u}.

We shortly discussed the numerical impact of our results on various
observables in the region $0.05 \le \s \le 0.25$, which is known to be free of
resonances. As an example,  we found the improvement on the forward-backward
asymmetry $\bar{A}_{\rm FB}(\s)$ defined in \Eref{eq:fbasym} to be striking:
the NNLL result is almost free of uncertainties due to the $\mu$-dependence.

\section*{Acknowledgement}
This work is partially supported by: the Swiss National Foundation; 
RTN, BBW-Contract No.~01.0357 and EC-Contract HPRN-CT-2002-00311 (EURIDICE);
NFSAT-PH 095-02 (CRDF 12050); SCOPES 7AMPJ062165.

%
%
\appendix
\section{Calculation techniques}
\label{section:techniques}
\subsection{Reducing tensor integrals with dimension-shifting techniques}
\label{section:tensor}
We follow Ref.~\cite{Tarasov} and derive a method that allows to express
tensor integrals in $D$ dimensions in terms of scalar integrals of
higher dimensions.

An arbitrary $L$ loop tensor integral with $N$ internal and $E$ external lines
can be written as a linear combination of integrals of the form (suppressing
Lorentz indices of $G^{(D)}$)
\begin{equation}
    \label{eq:Gd}
    G^{(D)}\Big(\big\{s_u\big\},\big\{m_v^2\big\}\Big) =
        \int\left(\prod\limits_{i=1}^{L} \frac{d^D k_i}{\left(2\,\pi\right)^D}
        \right) \prod\limits_{j=1}^{N}\, P_{\bar k_j,m_j}^{\nu_j}
        \prod\limits_{l=1}^{n_j} k_j^{\mu_{jl}}  \, ,
\end{equation}
where
\[
    P_{k,m}^{\nu} = \frac{1}{\big( k^2-m^2+i\,\epsilon \big)^\nu}
    \quad\text{and}\quad \bar k_j = \sum\limits_{n=1}^{L} \omega_{jn} \, k_n +
    \sum\limits_{m=1}^{E} \eta_{jm} \, q_m \,.
\]
$k_i$ and $q_j$ denote the loop and external momenta, respectively. The
matrices of incidences of the diagram, $\omega$ and $\eta$, have matrix
elements $\omega_{ij}, \eta_{ij} \in \{-1,0,1\}$. The quantities $\{s_u\}$ and
$\{m_v^2\}$ denote a set of scalar invariants formed from the external momenta
$q_j$ and a set of squared masses of the internal particles,
respectively. Generically, the exponents $\nu_i$ are equal to 1.  However,
often two or more internal lines are equipped with the same propagator. This
may be taken into account by reducing $N$ to $N^\eff < N$, thus increasing
some of the exponents $\nu_i$.

Applying the integral representations
\begin{align}
    \label{eq:tensor1}
    \frac{1}{\big( k^2-m^2+i\,\epsilon \big)^\nu} = &\,
    \frac{(-i)^\nu}{\Gamma(\nu)} \int\limits_0^\infty \!d\alpha \,
    \alpha^{\nu-1} \exp\Big[i\,\alpha \left(k^2-m^2+i\,\epsilon\right)\! \Big]
    \\ \intertext{and}
    \label{eq:tensor2}
    \prod\limits_{l=1}^{n_j} k_j^{\mu_{jl}} = &\,
    (-i)^{n_j}\prod\limits_{l=1}^{n_j} \frac{\partial}{\partial
    (a_j)_{\mu_{jl}}}\,\exp\!\Big[i(a_j\,k_j)\Big] \Bigg|_{a_j=0}
\end{align}
allows us to easily perform the integration over the loop momenta by using the
$D$ dimensional Gaussian integration formula
\[
    \int\!d^D k\,\exp\!\Big[i \left(A\,k^2 + 2(p\,k)\right)\! \Big] = i
        \left(\frac{\pi}{i\,A}\right)^{\frac{D}{2}}\, \exp\!
        \left[-\frac{i\,p^2}{A} \right].
\]
We find the following parametric representation:
\begin{multline}
    \label{eq:parametric}
    G^{(D)} = i^L \left( \frac{1}{4\,i\pi} \right)^{\!\frac{D\,L}{2}}
        \prod\limits_{j=1}^{N}\, \frac{(-i)^{n_j+\nu_j}}{\Gamma(\nu_j)} \times
        \prod\limits_{l=1}^{n_j}\, \frac{\partial}{\partial (a_j)_{\mu_{jl}}}
        \\ \times \int\limits_0^\infty \ldots \int\limits_0^\infty
        \frac{d\alpha_j\,\alpha_j^{\nu_j-1}}{\big[D(\alpha)\big]^\frac{D}{2}}\
        \exp\!\left[\frac{i\,Q\big(\{\bar s_i\},\alpha\big)}{D(\alpha)}
        -i\sum\limits_{r=1}^{N} \alpha_r \left(m_r^2-i\,\epsilon \right)
        \right] \Bigg|_{a_j=0}\,\,.
\end{multline}
The quantities $\bar s_i$ are scalar invariants involving the external momenta
$q_i$ and the auxiliary momenta $a_i$. $D(\alpha)$ arises from the integral
representations of the propagators: let $\vec{k}$ be the $L$-dimensional
vector that consists of all four-momentum loop vectors. The product of all
$P_{\bar k_j,m_j}^{\nu_j}$ can then be written as
\begin{eqnarray*}
  \prod\limits_{j=1}^{N}\,P_{\bar k_j,m_j}^{\nu_j} &=& \int\limits_0^\infty\,
  \left( \prod\limits_{j=1}^{N}\,d\alpha_i \right) f(\alpha) \exp\!\Big[i
  \left(\vec{k}^TB\vec{k} + (\vec{b}\,\vec{k}) +c\right) \Big],
\end{eqnarray*}
with $k_i$-independent quantities $f(\alpha),\,B,\,\vec{b}$ and $c$.
$D(\alpha)$ denotes the determinant of the $L\times{}L$ matrix $B$.

The differentiation of $G^{(D)}$ in \Eref{eq:parametric} with respect to
$a_j$ generates products of external momenta, metric tensors $g_{\mu\nu}$ and
polynomials $R(\alpha)$ and provides an additional factor
$D(\alpha)^{-1}$. Because of
\[
    R(\alpha) \exp\!\left[ -i \sum\limits_{r=1}^{N} \alpha_r m_r^2 \right] =
    R(i \partial) \exp\!\left[ -i \sum\limits_{r=1}^{N} \alpha_r m_r^2
    \right], \quad\text{with}\quad \partial_j = \frac{\partial}{\partial
    m_j^2}\, ,
\]
we may replace the polynomials $R(\alpha)$ with $R(i \partial)$. The
additional factor of $1/D(\alpha)$ can be absorbed by a redefinition of $D$,
i.e. by shifting $D$ to $D+2$ and multiplying with a factor $(4\,i\pi)^L$. The
crucial point is that all factors generated by differentiation with
respect to $a_j$ may be written as operators which do not depend on the
integral representations we have introduced in Eqs.~(\ref{eq:tensor1}),
(\ref{eq:tensor2}). Therefore, it is possible to write tensor integrals
in momentum space in terms of scalar ones without direct appeal to the
parametric representation (\ref{eq:parametric}):
\begin{equation}
    \label{eq:tensoreq}
    \int \left(\prod\limits_{i=1}^{L} \frac{d^D k_i}{\left(2\,\pi\right)^D}
    \right) \prod\limits_{j=1}^{N}\, P_{\bar k_j,m_j}^{\nu_j}\,
    \prod\limits_{l=1}^{n_j} k_j^{\mu_{jl}} = T\big(q,\partial,\bm{d^+} \big)
    \int\left(\prod\limits_{i=1}^{L} \frac{d^D k_i}{\left(2\,\pi\right)^D}
    \right) \, \prod\limits_{j=1}^N P_{\bar k_j,m_j}^{\nu_j}\, ,
\end{equation}
where the tensor operator $T$ (suppressing its Lorentz indices) is given by
\begin{multline}
    \label{eq:tensoroperator}
    T\big(q,\partial,\bm{d^+} \big) = \exp\!\Big[-i\, Q\big(\{\bar
    s_i\},\alpha\big) \left(4\,i\pi\right)^L \bm{d^+}\Big]\, \\ \times
    \prod\limits_{j=1}^{N}\, \prod\limits_{l=1}^{n_j} \frac{\partial}{\partial
    (a_j)_{\mu_{jl}}} \exp\!\Big[i\, Q\big(\{\bar s_i\},\alpha\big)
    \left(4\,i\pi\right)^L \bm{d^+}\Big] \Bigg|_{\substack{a_j=0 \\ \alpha_j =
    i \partial_j}}.
\end{multline}
The operator $\bm{d^+}$ shifts the space-time dimension of the integral by two
units:
\[
    \bm{d^+}\, G^{(D)}\Big(\big\{\bar s_i\big\},\big\{m_j^2\big\}\Big) =
    G^{(D+2)}\Big(\big\{\bar s_i\big\},\big\{m_j^2\big\}\Big).
\]
  Notice that throughout the derivation of
the tensor operator $T$ the masses $m_j$ must be kept as different
parameters. They are set to their original values only in the very end.

%
%
\subsection{Integration by parts}
\label{section:intbyparts}
According to general rules of $D$ dimensional integration, integrals of the
form
\[
    \int\!d^D k_i \,\frac{\partial}{\partial k_i^\mu}\, \frac{k_l^\mu}{\,
    \prod_{j=1}^N \Big(\bar k_j^2 - m_j^2 + i\,\epsilon\Big)^{\nu_j}}
\]
vanish. There may exist suitable linear combinations
\[
    \int\!d^D k_i \,\frac{\partial}{\partial k_i^\mu}\, \frac{\sum_l c_l 
    k_l^\mu+\sum_e d_e q_e^\mu}{\, \prod_{j=1}^N \Big(\bar k_j^2 - m_j^2 +
    i\,\epsilon\Big)^{\nu_j}}
\]
that lead to recurrence relations connecting the original integral to simpler
ones. The task of finding such recurrence relations, however, is in general a
nontrivial one. A criterion for irreducibility of multi-loop Feynman integrals
is presented in \cite{baikov}. In \cite{Tarasov}, the method of partial
integration is combined directly with the technique of reducing tensor
integrals by means of shifting the space-time dimension. \\ The integral
\begin{multline}
    F^{(D)}_{\nu_1\nu_2\nu_3\nu_4\nu_5} = \int\! d^D l\, d^D r\,
    I_{\nu_1\nu_2\nu_3\nu_4\nu_5} = \\ \int\! d^D l\, d^D r\,
    \frac{1}{\big[l^2\big]^{\nu_1} \big[r^2\big]^{\nu_2}
    \big[(l+r)^2\big]^{\nu_3} \big[(l+q)^2\big]^{\nu_4}
    \big[(r+p)^2-m_b^2\big]^{\nu_5}}
    \label{eq:integral}
\end{multline}
enters the calculation of diagrams \ref{fig:O1O2}d). At the
same time it is a very good example to illustrate the integration by parts
method. The operators $\bm{1^\pm}$, $\bm{2^\pm}$,... are defined through
\[
    \bm{1^\pm}I_{\nu_1\nu_2\nu_3\nu_4\nu_5} = I_{\nu_1\pm
    1\,\nu_2\nu_3\nu_4\nu_5},~\ldots\,.
\]
The present case is especially simple because we only need to calculate one
derivative. Using the shorthand notation $I_{\nu_1\nu_2\nu_3\nu_4\nu_5} =
I_{\{\nu_i\}}$ we get (for $\nu_i > 0 \, \forall \, i$):
\[
    \frac{\partial}{\partial r^\mu} \, r^\mu\, I_{\{\nu_i\}} = \Big[D -
    2\,\nu_2\,r^2\,\bm{2^+} - 2\,\nu_3\,r(l+r)\,\bm{3^+} -
    2\,\nu_5\,r(r+p)\,\bm{5^+} \Big] I_{\{\nu_i\}}.
\]
Scalar products of the form $(a\,b)$ we write as
$\left[a^2+b^2-(a-b)^2\right]/2$ and find
\[
    \frac{\partial}{\partial r^\mu} \, r^\mu\, I_{\{\nu_i\}} = \Big[D -
    2\,\nu_2 - \nu_3 - \nu_5 - \nu_3(\bm{2^-}-\bm{1^-})\,\bm{3^+} -
    \nu_5\,\bm{2^-}\,\bm{5^+} \Big] I_{\{\nu_i\}}.
\]
At this stage we might also reduce some of the scalar products by shifting the
dimension. The corresponding procedure is presented e.g. in
\cite{Tarasov}. In the present case, however, the pure integration by
parts approach suffices.  The identity
\[
    \int\!d^D r\, \frac{\partial}{\partial r_\mu} \, r^\mu\, I_{\{\nu_i\}}
    \equiv 0
\]
yields directly the desired recurrence relation for the integral
$F^{(D)}_{\nu_1\nu_2\nu_3\nu_4\nu_5}$:
\begin{equation}
    \label{eq:recrel}
    F^{(D)}_{\nu_1\nu_2\nu_3\nu_4\nu_5} =
    \frac{\nu_3(\bm{2^-}-\bm{1^-})\,\bm{3^+} + \nu_5\,\bm{2^-}\,\bm{5^+}}{D -
    2\,\nu_2 - \nu_3 - \nu_5} \, F^{(D)}_{\nu_1\nu_2\nu_3\nu_4\nu_5}.
\end{equation}
Subsequent application of this relation allows to express any integral
$F^{(D)}{\{\nu_i\}}$ with indices $\nu_i \in \bm{\mathbb{N}}^+$ as a sum over
integrals $F^{(D)}{\{\nu_i\}}$ with at least $\nu_1=0$ or $\nu_2=0$. 

The general procedure is the following:
\begin{itemize}
    \item One expresses suitable scalar products in the numerator of a given
    Feynman integrand in terms of inverse propagators $P_{\bar k,m}$ and
    cancels them down. It is important to notice that it is not always the best
    strategy to try to cancel down as many scalar products as possible. The
    resulting set of integrals to calculate highly depends on which scalar
    products one cancels down. The best way is to try a couple of different
    cancellation schemes and compare the resulting integrals.
    \item One writes the integral as a sum over tensor integrals of the form
    (\ref{eq:Gd}) with products of $k_i^\mu$. For each of those
    integrals the tensor operator $T$ is determined in order to reduce the
    problem to scalar integrals with shifted space-time dimension.
    \item One applies appropriate recurrence relations to reduce the number of
    propagators in the integrals, hoping to be able to solve the remaining
    integrals.
\end{itemize}

\section{Calculation of the diagrams \ref{fig:O1O2}\symbol{100}) } 
\label{section:diagram_d}
The contribution of the sum of diagrams \ref{fig:O1O2}d) is given by a
combination of integrals of the form
\begin{align}
  \int\! d^D l\, d^D r\, \frac{\prod_{i=1}^{n_l} l^{\mu_i} \prod_{j=1}^{n_r}
  r^{\rho_j}} {\big[l^2\big]^{\nu_1} \big[r^2\big]^{\nu_2}
  \big[(l+r)^2\big]^{\nu_3} \big[(l+q)^2\big]^{\nu_4}
  \big[(r+p)^2-m_b^2\big]^{\nu_5}} \,.
\end{align}
In this section we show how to solve these integrals with the methods
 presented in Appendices~\ref{section:tensor} and \ref{section:intbyparts}.
 The function $D(\a)$, which is independent of $n_l$ and $n_r$, is not needed
 in order to find the tensor operators $T$. Nevertheless, we give it as an
 illustration:
\begin{align*}
  D(\a) = (\a_1+\a_3+\a_4)\,(\a_2+\a_3+\a_5) - \a_3^2.
\end{align*}
The function $Q\big(\{\bar s_i\},\alpha\big)$, however, must be calculated
for each type of tensor integral. As an example we give $Q(\{\bar
s_i\},\alpha\}$ for $n_l=0$, $n_r=1$:
\begin{equation*}
  Q\big(\{\bar s_i\},\alpha\big) = -\left( \a_1 + \a_3 + \a_4 \right)\a_5
  \,(a_1 p) - \a_3\a_4 \,(a_1 q) - \frac{1}{4}\left(\a_1 + \a_3 +
  \a_4 \right) a_1^2.
\end{equation*}
The corresponding tensor operator  $T$ reads:
\begin{equation*}
  T^{\rho_1}(q,p,\partial,\bm{d^+})=16\p^2 \bm{d^+} \bigg[ q^{\rho_1} \pd_3 \pd_4+ p^{\rho_1} \pd_5
  \left(\pd_1 + \pd_3 + \pd_4 \right) \bigg].
\end{equation*}
The action of an operator $\partial_i$ on the integral $F^{(D)}_{\{\nu\}}$ is
\begin{equation}
    \partial_1^n \,F^{(D)}_{\nu_1\nu_2\nu_3\nu_4\nu_5} =
    \frac{\Gamma(\nu_1+n)}{\Gamma(\nu_1)}\,
    F^{(D)}_{\nu_1+n\,\nu_2\nu_3\nu_4\nu_5}\,,\dots\,.
\end{equation}
The next step is to repeatedly apply the recurrence relation (\ref{eq:recrel})
on the integrals $F^{(D)}_{\nu_1\nu_2\nu_3\nu_4\nu_5}$ until $\nu_1$ or
$\nu_2$ becomes zero. The problem is then reduced to the calculation of the
two types of integrals
\begin{equation}
  \label{eq:types}
  F^{(D)}_{0\nu_2\nu_3\nu_4\nu_5} \quad \text{and} \quad F^{(D)}_{\nu_1
  0\nu_3\nu_4\nu_5}.
\end{equation}
In the present calculation $D$ may take the values
\begin{equation}
    D = 4-2\epsilon,\,6-2\epsilon,\,8-2\epsilon \quad \text{or} \quad
    10-2\epsilon.
\end{equation}
It is important to note here that the denominator in \Eref{eq:recrel} can
become proportional to $\e$ for certain  values of $D,\,\n_2,\,\n_3$ and
$\n_5$. Thus, some of the integrals in (\ref{eq:types}) need to be calculated
up to $\ordo{\e^1}$. 

The first type of integrals ($F^{(D)}_{0\nu_2\nu_3\nu_4\nu_5}$) can easily be
solved individually by using a single Mellin-Barnes approach. This method
naturally results in an expansion in $\s$. Furthermore, the
occasionally needed $\ordo{\e^1}$ terms are easily obtained since the
expansion in $\e$ is done only in the very end. We now turn to the much more
complicated calculation of the second set of integrals. Instead of calculating
every single occurring integral individually, we derive a general formula for
$F^{(D)}_{\nu_10\nu_3\nu_4\nu_5}$ where we are left with a three-dimensional
Feynman parameter integral:
\begin{eqnarray}
  \label{eq:generic2}
  F^{(D)}_{\nu_10\nu_3\nu_4\nu_5} & = & (-1)^{\nu_1 + \nu_3
  \nu_4 + \nu_5 + D} \frac{\Gamma(\nu_1+\nu_3+\nu_4+\nu_5)}
  {\Gamma(\nu_1) \, \Gamma(\nu_3) \, \Gamma(\nu_4) \, \Gamma(\nu_5)}
  \int\limits_0^1 \!d u \, d x \, d y \, u^{\nu_1+\nu_3+\nu_4-1-D/2} \nn \\ &&
  \times (1-u)^{D/2-\nu_3-1} \, x^{\nu_3 -1} \, (1-x)^{\nu_3 + \nu_5 -1
  -D/2} \, y^{\nu_1-1} \, (1-y)^{D/2 -1 -\nu_1} \nn \\ && \times
  \hat{\Delta}^{D-\nu_1-\nu_3-\nu_4-\nu_5} \\ \hat{\Delta} &=& m_b^2 \,
  (1-x)\, (1-u\,y) - s\,x\,y\,u - i\delta. \nn
\end{eqnarray}
We now replace all occurrences of $F^{(D)}_{\nu_10\nu_3\nu_4\nu_5}$ according
to \Eref{eq:generic2} and are left with a three-dimensional integral over
a rather lengthy integrand. This integrand can be split up into three
different parts:
\begin{itemize}
\item A part with no additional divergences arising from the integrations.
\item A part with problematic $x$-integration.
\item A part with problematic $u$-integration.
\end{itemize}
In the first part, the regulator $\epsilon$ is not needed at all and may be
set equal to zero at the very beginning. The occurring integrals can then
either be performed directly or with the use of a single Mellin-Barnes
representation.  The second part boils down to two different integrals, which
can both be computed with subtraction methods. The last part is clearly the
most difficult one. It can be reduced to three integrals which we calculate
using a double Mellin-Barnes representation. Since this double Mellin-Barnes
is very different from the one presented in Subsection 3.1.4 in \cite{AAGW},
we give, as an example, the needed procedure to calculate one of the three
integrals. Specifically, we have to deal with the integrals
\begin{eqnarray}
  \label{eq:u_example}
  I_{j} & = & \int\limits_0^1\!dx\,\int\limits_0^1\!dy\,\int\limits_0^1\!du\,
  \frac{u^{\e}\,(1-y)^{2-\e}\,x^j\,(1-x)^\e}{(1-u)^{1+\e}\,
  \left((1-x)(1-u\,y) - \s\,x\,u\,y -i\d\right)^{1+2\e}},
\end{eqnarray}
where $j$ can take the values $0,\,1$ and $2$. We focus on the case where
$j=0$. We introduce a first Mellin-Barnes integral in the complex $t$-plane
with the identifications (for notation see e.g.~\cite{AAGW}):
\begin{equation*}
  K^2 \leftrightarrow (1-x)(1-u\,y), \quad M^2 \leftrightarrow \s\,x\,u\,y+i\d,
  \quad \l=1+2\e,
\end{equation*}
and get
\begin{eqnarray}
  \label{eq:mb1}
  I_0 & = &
  \frac{\text{e}^{-i\pi(1+2\e)}}{2\,i\p\,\Gamma(1+2\e)}\int\limits_\g \!
  dt\,\int\limits_0^1\!dx\,\int\limits_0^1\!dy\,\int\limits_0^1\!du\,
  \Gamma(-t) \Gamma(1+2\e+t) \nn \\ && \times
  \frac{u^{t+\e}y^t(1-y)^{2-\e}x^t}{(1-u)^{1+\e}
  (1-x)^{1+\e+t}(1-u\,y)^{1+2\e+t}}\,\s^t. 
\end{eqnarray}
The path $\g$ lies in the left half-plane and can be chosen arbitrarily close
to the imaginary $t$-axis. We introduce a second Mellin-Barnes representation
in the complex $t'$-plane for the last factor in the denominator of
\Eref{eq:mb1}. For this, we rewrite $1-u\,y$ as $1-u+u(1-y)$ and make the
following identifications:
\begin{equation*}
  K^2 \leftrightarrow u(1-y), \quad M^2 \leftrightarrow -(1-u),
  \quad \l=1+2\e+t,
\end{equation*}
yielding
\begin{eqnarray}
  \label{eq:mb2}
  I_0 & = &
  \frac{\text{e}^{-i\pi(1+2\e)}}{\left(2\,i\p\right)^2\,\Gamma(1+2\e)}
  \int\limits_{\g'} \! dt'\,\int\limits_\g\!
  dt\,\int\limits_0^1\!dx\,\int\limits_0^1\!dy\,\int\limits_0^1\!du\,
  \Gamma(-t) \Gamma(-t') \nn \\ &&  \times
  \Gamma(1+2\e+t+t')\,\frac{y^t(1-y)^{1-3\e-t-t'}x^t}
  {u^{1+\e+t'}(1-u)^{1+\e-t'}(1-x)^{1+\e+t}}\,\s^t.
\end{eqnarray}
The path $\g'$ lies to the left of the imaginary $t'$-axis and can again be
chosen arbitrarily close to that axis.  The parameter integrals can now be
performed and give products of Euler Beta-functions. We work out the remaining
integrals over $t$ and $t'$ applying the residue theorem. For this, we close
the $t$-integral in the right half-plane and focus on the enclosed
poles. There are two different sequences of poles, namely poles that depend on
$t'$ (coupled poles) and poles that do not (uncoupled poles). The latter poles
lie at the following positions:
\begin{itemize}
\item $t = 0,\,1,\,2,\, \dots,\,S,\, \dots\,,$
\item $t = 0-\e,\,1-\e,\,2-\e,\, \dots,\,S-\e,\, \dots\,,$
\end{itemize}
Note here that $I_0$ exists only for negative values of $\e$. The pole located
at $t=-\e$ therefore lies in the right half-plane and needs to be taken into
account. Since we are interested in an expansion in $\s$, we can truncate the
two pole sequences at a suitable $S$.  After calculating the necessary
residues, we close the $t'$-integral in the right half-plane as well and are
arrive at pole sequences situated at the following positions:
\begin{itemize}
\item  $t' = 0,\,1,\,2,\, \dots\,,$
\item   $t' = 0-\e,\,1-\e,\,2-\e,\, \dots\,,$
\item   $t' = 2-N-3\e,\,3-N-3\e,\,4-N-3\e,\, \dots \quad \quad$ for $t=N,\quad
  N \in \mathbb N\,,$\\
  $t' = 2-N-2\e,\,3-N-2\e,\,4-N-2\e,\, \dots \quad \quad$ for $t=N-\e,\quad N
  \in \mathbb N\,.$
\end{itemize}
For $N\ge3$, some of the poles above lie in the left $t'$-half-plane and must
be omitted. Unlike the procedure given in Subsection 3.1.4 of \cite{AAGW}, we
need to sum up the residues of all poles in the enclosed area.

Calculating the contributions of the coupled poles in $t$, which lie at
$t=2+n-3\e-t', n \in \mathbb N$, yields an expression that is proportional to
$\s^{2+n-3\e-t'}$. Two problems now arise if one closes the integration path
of the $t'$-integral in the right half-plane: due to the $-t'$ in the exponent
of $\s$, one gets an expansion in inverse powers of $\s$, forcing one to
calculate the residues of all enclosed poles. The second problem is even
worse: for any given value of $n$, there always exists an infinite pole series
in $t'$ which contributes to the desired result. Thus, one also has to
consider the infinite pole series in $t$. In order to avoid these problems, we
close the integration path in the left half-plane of $t'$. The poles are then
located at
\begin{itemize}
\item $t' = -1-\e,\,-2-\e,\,-3-\e,\, \dots\,,$
\item $t' = -1-2\e,\,-2-2\e,\,-3-2\e,\, \dots\,,$
\item $t' = -1-3\e,\,-2-3\e,\,-3-3\e,\, \dots\,.$
\end{itemize}
After calculating the necessary residues we obtain the result for $I_0$.
The results for $I_1$ and $I_2$ are calculated in an analogous way.

\section{Solution of the Renormalization group equation for the Wilson
  coefficients} 
\label{appendix:wilson}
The Wilson coefficients satisfy the renormalization group equation
\begin{eqnarray}
  \label{eq:rge1}
  \frac{d}{d\ln \m} \vec{C}(\m) & = & \gamma^T(\alphas)\,\vec{C}(\m)\,,
\end{eqnarray}
where $\gamma(\alphas)$ is the anomalous dimension matrix. This matrix
can be written as a Taylor series in $\alphas$:
\begin{eqnarray*}
  \gamma(\alphas) &=& \gamma^{(0)} \alphasFourPi + \gamma^{(1)}
  \left(\alphasFourPi \right)^2 + \gamma^{(2)} \left( \alphasFourPi \right)^3
  + \ldots\,.
\end{eqnarray*}
The general solution of \Eref{eq:rge1} can be expressed with the
evolution matrix $U(\mu,\mu_0)$:
\begin{eqnarray}
  \label{eq:evolution}
   \vec{C}(\mu) &=& U(\mu,\mu_0)\,\vec{C}(\mu_0), \nn \\ U(\mu_0,\mu_0) &=& 1.
\end{eqnarray}
The aim in this section is to find a handy expression for $U(\m,\m_0)$.\\
The matrix $\gamma^{(0)}$ can be diagonalized. We introduce new quantities in
the following way:
\begin{eqnarray}
  \label{eq:tilde}
  \vec{C}(\mu) &=& V\,\vec{\widetilde{C}}(\mu), \nn \\ \gamma^{(i)} &=&
  V\,\widetilde{\gamma}^{(i)}\,V^{-1}\,,\,\,\, i=0,1,2,\ldots \,,\\
  U(\mu,\mu_0) &=& V\,\widetilde{U}(\mu,\mu_0)\,V^{-1}. \nn
\end{eqnarray}
The matrix $V$ is chosen such that $\widetilde{\gamma}^{(0)}$ is diagonal. One
can check that the new quantities satisfy equations similar to 
(\ref{eq:rge1}) and (\ref{eq:evolution}):
\begin{eqnarray}
  \label{eq:rge2}
  \frac{d}{d\ln \m} \vec{\widetilde{C}}(\m) & = &
  \widetilde{\gamma}^T(\alphas)\,\vec{\widetilde{C}}(\m), \nn \\
  \vec{\widetilde{C}}(\mu) &=&
  \widetilde{U}(\mu,\mu_0)\,\vec{\widetilde{C}}(\mu_0), \\
  \widetilde{U}(\mu_0,\mu_0) &=& 1.  \nn
\end{eqnarray}
We will now construct a solution to \Eref{eq:rge2}. Once this solution is
found, we can easily gain the solution of the initial problem for the
non-diagonal $\gamma^{(0)}$. The evolution matrix $\widetilde{U}(\mu,\mu_0)$
satisfies the same equation as $\vec{\widetilde{C}}(\mu)$ itself:
\begin{eqnarray}
  \label{eq:umatrix}
  \frac{d}{d\ln \m} \widetilde{U}(\m,\mu_0) & = &
  \widetilde{\gamma}^T(\alphas) \,\widetilde{U}(\m,\mu_0).
\end{eqnarray}
We make the following ansatz for $\widetilde{U}(\m,\mu_0)$:
\begin{eqnarray}
  \label{eq:ansatz}
  \widetilde{U}(\m,\mu_0) &=& \left(1+\sum_{i=1}^{\infty} \left(
  \frac{\alphas(\mu)}{4\pi} \right)^i \widetilde{J}_{i}
  \right)\,\widetilde{U}^{(0)}(\mu,\mu_0)\,\widetilde{K},
\end{eqnarray}
where $\widetilde{U}^{(0)}(\mu,\mu_0)$ solves \Eref{eq:umatrix} to
leading logarithmic approximation and is given by
\begin{eqnarray*}
  \widetilde{U}^{(0)}(\mu,\mu_0) &=& \left(
  \left(\frac{\alphas(\mu_0)}{\alphas(\mu)}
  \right)^{\frac{\vec{\gamma}^{(0)}}{2\b_0}} \right)_D.
\end{eqnarray*}
The vector $\vec{\gamma}^{(0)}$ collects the diagonal elements of
$\widetilde{\gamma}^{(0)}$. The matrix $\widetilde{K}$ must be chosen such
that the boundary condition given in \Eref{eq:rge2} is met. The
quantities $\b_i,\,\,\, i=0,1,2,\ldots$ appear in the RGE for $\alphas$:
\begin{eqnarray*}
\frac{d}{d\ln \m} \alphas(\m) &=& -2 \sum_{i=0}^{\infty}
\frac{\alphas(\m)^{i+2}}{\left(4\pi\right)^{i+1}} \b_i.
\end{eqnarray*}
Inserting the ansatz (\ref{eq:ansatz}) into \Eref{eq:umatrix} and using the
explicit expression for $\widetilde{U}^{(0)}(\mu,\mu_0)$, the lhs and the rhs
of this equation can be written as
\begin{eqnarray*}
  \mbox{lhs}=\sum_{j=1}^{\infty}\,\left( \frac{\alphas(\mu)}{4\pi} \right)^j
  L_{j}\,\widetilde{U}^{(0)}(\mu,\mu_0)\,\widetilde{K}, \\
  \mbox{rhs}=\sum_{j=1}^{\infty}\,\left( \frac{\alphas(\mu)}{4\pi} \right)^j
  R_{j}\,\widetilde{U}^{(0)}(\mu,\mu_0)\,\widetilde{K}.
\end{eqnarray*}
The unknown matrices $\widetilde{J}_i$ can now be constructed order by order
in $\alphas$ through the relations $L_j=R_j$. We give the explicit solutions
to $\widetilde{J}_1$ and $\widetilde{J}_2$ since we need them to find the
Wilson coefficients $C_i(\m)$ to NNLL precision:
\begin{eqnarray}
  \label{eq:jsol}
  \widetilde{J}_{1,ij} &=& \d_{ij} \vec{\gamma}_i^{(0)} \frac{\b_1}{2\b_0^2} -
  \frac{\widetilde{\gamma}_{ij}^{(1)}\raisebox{1.7ex}{\scriptsize{T}}}{2\b_0+
  \vec{\gamma}_i^{(0)} - \vec{\gamma}_j^{(0)}}, \\ 
  \label{eq:jsol2}
  \widetilde{J}_{2,ij} &=&
  \d_{ij} \vec{\gamma}_i^{(0)} \frac{\b_2}{4\b_0^2} -
  \frac{\widetilde{\gamma}_{ij}^{(2)}\raisebox{1.7ex}{\scriptsize{T}} +
  \left(2\b_1-\tfrac{\b_1}{\b_0}\vec{\gamma}_j^{(0)}
  \right)\widetilde{J}_{1,ij}+
  \left(\widetilde{\gamma}^{(1)}\raisebox{1.7ex}{\scriptsize{T}}\,
  \widetilde{J}_1 \right)_{ij}} {4\b_0+\vec{\gamma}_i^{(0)} -
  \vec{\gamma}_j^{(0)}}.
\end{eqnarray}
The result for $\widetilde{J}_1$ agrees with the one given in Section III of
\cite{Buchalla:1995vs}. After we did the calculation for $\widetilde{J}_2$, we
found out that the result already exists in the literature
\cite{Beneke:2001at}. The two results agree as well.\\
The matrix $\widetilde{K}$ is given through
\begin{eqnarray}
  \label{eq:constant}
  \widetilde{K} &=& 1 - \frac{\alphas(\mu_0)}{4\pi}\widetilde{J}_1 -
  \left(\frac{\alphas(\mu_0)}{4\pi}\right)^2 \left(\widetilde{J}_2 -
  \widetilde{J}_1^2 \right) + \ordo{\alphas^3}.
\end{eqnarray}
With these informations at hand, we can present the evolution matrix for the
initial problem given in Eqs.~(\ref{eq:rge1}) and (\ref{eq:evolution}):
\begin{eqnarray}
  \label{eq:ufinal}
  U(\mu,\mu_0) &=& V\,\left(1+\frac{\alphas(\mu)}{4\pi}\,\widetilde{J}_1 +
  \left( \frac{\alphas(\mu)}{4\pi}\right)^2\,\widetilde{J}_2 \right)
  \widetilde{U}^{(0)}(\mu,\mu_0)\,\widetilde{K}\,V^{-1} + \ordo{\alphas^3}.
\end{eqnarray}

\section{One-Loop Matrix Elements of the Four-Quark Operators}
\label{appendix:oneloop} In order to fix the counterterms $F_{i,u \to
  \rm{4 quark}}^{\rm{ct}(7,9)}$ ($i=1,2$) in
Eq.~(\ref{fi4quark}), we need the one-loop matrix elements $\bra d\,
\ell^+ \ell^-|O_j|b \ket_{\text{1-loop}}$ of the four-quark operators $O_1^u$,
$O_2^u$, $O_4$, $O_{11}^u$ and $O_{12}^u$. Due to the $1/\epsilon$ factor in
Eq.~(\ref{fi4quark}) they are needed up to $\ordo{\epsilon^1}$. The
explicit results read
\begin{align*}
    \bra d\,\ell^+\ell^- |O_2^u|b\ket_{\text{1-loop}} = & \,\,\left(
      \frac{\mu}{m_b}\right)^{2\epsilon} \Bigg\{\frac{4}{9\,\epsilon} +
    \frac{4}{27} \Big[\, 2 + 3\,i\pi - 3\,L_s\, \Big] + \\ &
    \frac{\epsilon}{81} \Big[\, 52+24\,i\pi - 21\,\pi^2 - (24+36\,i\pi)
    L_s + 18\,L_s^2\, \Big]\! \Bigg\}\, \bra
    \widetilde{O}_9\ket_{\text{tree}}\, , \\ \\
    \bra d\,\ell^+\ell^- |O_1^u|b\ket_{\text{1-loop}} = &\, \frac{4}{3}\, \bra
    d\,\ell^+\ell^- |O_2^u|b\ket_{\text{1-loop}} \, , \\ \\
    \bra d\,\ell^+\ell^- |O_4|b\ket_{\text{1-loop}} = &
    -\left(\frac{\mu}{m_b}\right)^{2\epsilon} \Bigg\{\! \left[\frac{4}{9} +
      \frac{\epsilon}{945} \left(70\, \s + 7\, \s^2 +
        \s^3\right)\right] \, \bra \widetilde{O}_7\ket_{\text{tree}} \\ & +
    \Bigg[ \frac{16}{27\,\epsilon} + \frac{2}{8505} \left( - 420 + 1260\, i\pi
      - 1260\, L_s + 252\, \s + 27\, \s^2 + 4\, \s^3 \right) \\ &
    + \frac{4\,\epsilon }{8505} \left(420\, i\pi + 910 - 630\,L_s\,i\pi -
      420\, L_s - 315\,\pi^2 \right. \\ & + \left. \left. 315\,L_s^2 -
        126\,\s+ \s^3 \right)\Bigg]\bra
    \widetilde{O}_9\ket_{\text{tree}}\right\} , \\
\end{align*}
\begin{align*}
    \bra d\,\ell^+\ell^- |O_{11}^u|b\ket_{\text{1-loop}} = & -\frac{64}{27}
    \left(\frac{\mu}{m_b}\right )^{2\epsilon} \left(1 +
      \frac{5}{3}\,\epsilon + i\pi\,\epsilon - L_s\,\epsilon \right) \, \bra
    \widetilde{O}_9\ket_{\text{tree}} \, , \\ \\
    \bra d\,\ell^+\ell^- |O_{12}^u|b\ket_{\text{1-loop}} = &\, \frac{3}{4}\,
    \bra d\,\ell^+\ell^- |O_{11}^u|b\ket_{\text{1-loop}} \, . \\
\end{align*}

%
%
\section{Finite Bremsstrahlung Corrections}
\label{appendix:brems} 
In \Sref{section:decay_width} those bremsstrahlung contributions were taken
into account which generate infrared and collinear singularities. Combined
with virtual contributions which also suffer from such singularities, a finite
result was obtained. In this appendix we discuss the remaining finite
bremsstrahlung corrections which are encoded in the last two terms of
\Eref{rarewidth}. Being finite, these terms can be directly calculated in $d=4$
dimensions.

The sum of the bremsstrahlung contributions from $O_7-O_8$ and $O_8-O_9$
interference terms and the $O_8-O_8$ term can be written as
\begin{multline}
    \frac{d \Gamma^{\Brems,\,\text{A}}}{d\s} = \frac{d
    \Gamma_{78}^\Brems}{d\s}\, + \,\frac{d \Gamma_{89}^\Brems}{d\s}\, +
    \,\frac{d \Gamma_{88}^\Brems}{d\s}\,=\\
    \left(\frac{\alpha_{\text{em}}}{4\,\pi}\right)^2\,
    \left(\frac{\alpha_s}{4\,\pi} \right) \frac{m_{b,\pole}^5\,|\xi_t|^2
    \, G_F^2}{48\, \pi^3} \times \big( 2\, \text{Re} \left[c_{78}\,\tau_{78} +
    c_{89}\,\tau_{89}\right] + c_{88}\,\tau_{88} \big),
\end{multline}
where
\begin{equation}
    c_{78} = C_F \cdot \wtC_7^{(0,\eff)} \wtC_8^{(0,\eff)*}, \quad c_{89} =
    C_F \cdot \wtC_8^{(0,\eff)} \wtC_9^{(0,\eff)*} , \quad c_{88} = C_F \cdot
    \left| \wtC_8^{(0,\eff)} \right|^2.
\end{equation}
For the quantities $\tau_{78}$, $\tau_{89}$ and $\tau_{88}$ we refer to
\cite{Asa2}.

The remaining bremsstrahlung contributions all involve the diagrams with an
$O_{1,2}^u$ or $O_{1,2}^c$ insertion where the gluon is emitted from the $u$-
or $c$-quark loop, respectively. The corresponding bremsstrahlung matrix
elements depend on the functions $\bar\Delta i_{23}^{(u,c)}$, $\bar\Delta i_{27}^{(u,c)}$. In $d=4$ dimensions we find
\begin{align*}
    \label{eq:deltaik}
    \bar\Delta i_{23}^{(u)} &= 8\, (\qr) \int_0^1\! dx\,dy\, \frac{x\, y\,
    (1-y)^2}{C^{(u)}} \, , & \bar\Delta i_{23}^{(c)} &= 8\, (\qr) \int_0^1\!
    dx\,dy\, \frac{x\, y\, (1-y)^2}{C^{(c)}} \, , \\ \bar\Delta i_{27}^{(u)}
    &= 8\, (\qr) \int_0^1\! dx\,dy\, \frac{y\, (1-y)^2}{C^{(u)}} \, , &
    \bar\Delta i_{27}^{(c)} &= 8\, (\qr) \int_0^1\! dx\,dy\, \frac{y\,
    (1-y)^2}{C^{(c)}} \, ,
\end{align*}
where
\begin{alignat*}{2}
    C^{(u)} & = & - 2\, x\, y\, (1-y) (\qr) - q^2\, y\, (1-y) - i\,\delta \, ,
    \\ C^{(c)} & = m_c^2 & - 2\, x\, y\, (1-y) (\qr) - q^2\, y\, (1-y) -
    i\,\delta \, .
\end{alignat*}\\
\noindent The analytical expressions for $\bar\Delta i_{23}^{(c)}$ and
    $\bar\Delta i_{27}^{(c)}$ can be written in terms of functions $G_i(t)$:
\begin{align}
    \bar\Delta i_{23}^{(c)} & = -2 + \frac{4}{w-\s} \left[ z\,
        G_{-1}\!\left(\frac{\s}{z}\right) - z\,
        G_{-1}\!\left(\frac{w}{z}\right) - \frac{\s}{2}\,
        G_0\!\left(\frac{\s}{z}\right) + \frac{\s}{2}\,G_0\!\left(\frac{w}{z}
        \right) \right], \\ \nn \\ \bar\Delta i_{27}^{(c)} & = 2 \left[
        G_0\!\left(\frac{\s}{z}\right) - G_0\!\left(\frac{w}{z}\right)
        \right],
\end{align}
where $z=m_c^2/m_b^2$. $G_k(t)$ ($k \ge -1$) is defined through the integral
\[
    G_k(t) = \int\limits_0^1\! d x \, x^k \, \ln\!\big[
        1-t\,x(1-x)-i\,\delta\big], \quad G_1(t) = \frac{1}{2} G_0(t).
\]
Explicitly, the functions $G_{-1}(t)$ and $G_0(t)$ read
\begin{align}
    G_{-1}(t) & =
    \begin{cases}
        2\,\pi\, \arctan\!\left( \sqrt\frac{4-t}{t}\right) - \frac{\pi^2}{2} -
        2\,\arctan^2\!\left( \sqrt\frac{4-t}{t}\right), & t < 4 \\ \\ -
        2\,i\pi\,\ln\!\left( \frac{\sqrt{t}+\sqrt{t-4}}{2}\right) -
        \frac{\pi^2}{2} + 2\, \ln^2\!\left(
        \frac{\sqrt{t}+\sqrt{t-4}}{2}\right), & t > 4
    \end{cases}\, , \\ \nn \\
    G_0(t) & =
    \begin{cases}
        \pi \, \sqrt{\frac{4-t}{t}} - 2 - 2\, \sqrt{\frac{4-t}{t}}\,
            \arctan\!\left( \sqrt\frac{4-t}{t}\right)), & t < 4 \\ \\ -i\pi
            \sqrt{\frac{t-4}{t}} - 2 + 2\, \sqrt{\frac{t-4}{t}}\,\ln\!\left(
            \frac{\sqrt{t}+\sqrt{t-4}}{2}\right), & t > 4
    \end{cases}\, .
\end{align}
The quantities $\bar\Delta i_{j}^{(u)}$ we obtain from $\bar\Delta
i_{j}^{(c)}$ in the limit $z \to 0$:
\begin{align*}
    \bar\Delta i_{23}^{(u)} & = -2 + \frac{2\,\s}{w-\s} \big[ \ln(w)- \ln(\s)
    \big] \, , \\ \bar\Delta i_{27}^{(u)} & = -2 \big[ \ln(w)- \ln(\s) \big].
\end{align*}

Following \cite{Asa2}, we write
\begin{multline}
     \label{eq:tau}
     \frac{d \Gamma^{\Brems,\,\text{B}}}{d\s} = \left(
    \frac{\alpha_{\text{em}}}{4\,\pi} \right)^2 \left( \frac{\alpha_s}{4\,\pi}
    \right) \frac{G_F^2\, m_{b,\pole}^5 \,|\xi_t|^2}{48\,\pi^3} \times \\
    \int\limits_{\s}^{1}\!d w\, \Big\{\!  \left( c_{11} + c_{12} + c_{22}
    \right) \tau_{22} + 2\,\text{Re} \big[ \left( c_{17} + c_{27} \right)
    \tau_{27} + \left( c_{18} + c_{28} \right) \tau_{28} + \left( c_{19} +
    c_{29} \right) \tau_{29} \big] \Big\}.
\end{multline}
Expressed in terms of the quantities $\bar\Delta i_{23}^\eff$ and $\bar\Delta
i_{27}^\eff$, defined by
\begin{align}
    \bar\Delta i_{23}^\eff = & -\frac{\xi_u}{\xi_t} \bar\Delta
    i_{23}^{(u)} -\frac{\xi_c}{\xi_t} \bar\Delta i_{23}^{(c)} , \\
    \bar\Delta i_{27}^\eff = & -\frac{\xi_u}{\xi_t} \bar\Delta
    i_{27}^{(u)} -\frac{\xi_c}{\xi_t} \bar\Delta i_{27}^{(c)} ,
\end{align}
the quantities $\tau_{ij}$ introduced in \Eref{eq:tau} read
\begin{multline}
    \tau_{22} = \frac{8}{27} \frac{(w-\s)(1-w)^2}{\s\,w^3} \times \bigg\{\!
    \Big[ 3\,w^2 + 2\,\s^2(2+w) - \s\,w\,(5-2\,w) \Big] \left| \bar\Delta
    i_{23}^\eff \right|^2 + \\ \Big[ 2\,\s^2\,(2+\,w) + \s\,w\,(1+2\,w) \Big]
    \left| \bar\Delta i_{27}^\eff \right|^2 + 4\,\s \Big[w\,(1-w) - \s\,(2+w)
    \Big] \cdot \text{Re}\left[ \bar\Delta i_{23}^\eff \bar\Delta
    i_{27}^{\eff*} \right] \!\!\bigg\},
\end{multline}

\begin{multline}
    \tau_{27} = \frac{8}{3}\frac{1}{\s\,w} \times \bigg\{\!\Big[
    (1-w)\left(4\,\s^2 - \s\,w+w^2\right) + \s\,w\,(4+\s-w) \ln(w) \Big]
    \bar\Delta i_{23}^\eff \\ - \Big[ 4\,\s^2\,(1-w) + \s\,w\,(4 + \s
    -w)\,\ln(w) \Big] \bar\Delta i_{27}^\eff \bigg\},
\end{multline}

\begin{multline}
    \tau_{28} = \frac{8}{9}\frac{1}{\s\,w\,(w-\s)} \times \Bigg\{\!
    \Big[(w-s)^2(2\,\s-w)(1-w) \Big] \bar\Delta i_{23}^\eff - \Big[
    2\,\s\,(w-\s)^2(1-w) \Big] \bar\Delta i_{27}^\eff \\ + \s\,w \Big[
    (1+2\,\s-2\,w)\bar\Delta i_{23}^\eff - 2\, (1+\s-w) \bar\Delta
    i_{27}^\eff\Big] \cdot \ln\!\left[ \frac{\s}{(1+\s-w)(w^2+\s\,(1-w))}
    \right] \!\!\Bigg\},
\end{multline}

\begin{multline}
    \tau_{29} = \frac{4}{3}\frac{1}{w}\times \bigg\{\!  \Big[ 2\,\s(1-w)(\s+w)
    + 4\,\s\,w\ln(w) \Big] \bar\Delta i_{23}^\eff - \\ \Big[ 2\,\s(1-w)(\s+w)
    + w (3\,\s+w) \ln(w) \Big] \bar\Delta i_{27}^\eff \bigg\}.
\end{multline}
The coefficients $c_{ij}$ include the dependence on the Wilson coefficients
and the color factors.
\begin{align}
    c_{11} = & \, C_{\tau_1} \cdot \left|C_1^{(0)}\right|^2, & c_{17} = & \,
    C_{\tau_2} \cdot C_1^{(0)} \wtC_7^{(0,\eff)*}, & c_{27} = & \, C_F \cdot
    C_2^{(0)} \wtC_7^{(0,\eff)*}, \nn \\ c_{12} = & \, C_{\tau_2} \cdot
    2\,\text{Re}\!\left[ C_1^{(0)} C_2^{(0)*}\right], & c_{18} = & \,
    C_{\tau_2} \cdot C_1^{(0)} \wtC_8^{(0,\eff)*}, & c_{28} = & \, C_F \cdot
    C_2^{(0)} \wtC_8^{(0,\eff)*}, \\ c_{22} = & \, C_F \cdot
    \left|C_2^{(0)}\right|^2, & c_{19} = & \, C_{\tau_2} \cdot C_1^{(0)}
    \wtC_9^{(0,\eff)*}, & c_{29} = & \, C_F \cdot C_2^{(0)}
    \wtC_9^{(0,\eff)*}. \nn
\end{align}
The Wilson coefficients $C_{7,8,9,10}^{\rm eff}$ are given in \Eref{effcoeff}
and numerical values for the coefficients $C_i^{(0)}$ can be found in
Table~\ref{tableA}.  The color factors $C_F$, $C_{\tau_1}$ and $C_{\tau_2}$
are presented in \Sref{section:eff}.
 


\begin{thebibliography}{99}

\bibitem{CLEOold}
M.S.~Alam {\it et al.} (CLEO Collab.), \prl{74}{1995}{2885}.

\bibitem{CLEOmed}
S.~Ahmed {\it et al.} (CLEO Collab.) CLEO CONF 99-10, 
\hepex{9908022}.

\bibitem{CLEOnew}
S.~Chen {\it et al.} (CLEO Collab.), \prl{87}{2001}{251807}, \hepex{0108032}.

\bibitem{BRALEPH}
R.~Barate {\it et al.} (ALEPH Collab.), \plb{429}{1998}{169}.

\bibitem{BELLE01}
K.~Abe {\it et al.} (BELLE Collab.), \plb{511}{2001}{151},
\hepex{0103042}.

\bibitem{BABAR02}
B.~Aubert {\it et al.} (BABAR Collab.), \hepex{0207074} and \hepex{0207076}.

\bibitem{Jessop2003}
C.~Jessop, SLAC-PUB-9610.

\bibitem{BELLEbsll2} 
J.~Kaneko {\it et al.} [BELLE Collaboration],
\prl{90}{2003}{021801},
\hepex{0208029}.

\bibitem{BELLEbsll3} 
K.~Abe {\it et al.} [BELLE Collaboration],
\hepex{0107072}.

\bibitem{Babarbsll}
B.~Aubert {\it et al.}  [BABAR Collaboration],
\hepex{0308016}.

\bibitem{Ligeti:1996yz}
Z.~Ligeti and M.~B.~Wise,
\prdd {53}{1996}{4937}, \hepph{9512225}.

\bibitem{Falk:1994dh}
A.~F.~Falk, M.~Luke and M.~J.~Savage,
\prdd {49}{1994}{3367}, \hepph{9308288}.

\bibitem{Ali:1997bm}
A.~Ali, G.~Hiller, L.~T.~Handoko and T.~Morozumi,
\prdd {55}{1997}{4105}, \hepph{9609449}.

\bibitem{chen:1997} J-W.~Chen, G.~Rupak and M.~J.~Savage,
\plb {410}{1997}{285}, \hepph{9705219}.

\bibitem{Buchalla:1998ky}
G.~Buchalla, G.~Isidori and S.~J.~Rey,
\npb {511}{1998}{594}, \hepph{9705253}.

\bibitem{Buchalla:1998mt}
G.~Buchalla and G.~Isidori,
\npb {525}{1998}{333}, \hepph{9801456}.

\bibitem{Kruger:1996dt}
F.~Kruger and L.~M.~Sehgal,
\prdd{55}{1997}{2799}, \hepph{9608361}.

\bibitem{Ball}
A.~Ali, P.~Ball, L.T.~Handoko, G.~Hiller,
\prdd {61}{2000}{074024}, \hepph{9910221}.

\bibitem{Lunghi}
E.~Lunghi and I.~Scimemi,
\npb {574}{2000}{43}, \hepph{9912430}.

\bibitem{Silvestrini}
E.~Lunghi, A.~ Masiero, I.~Scimemi and L.~Silvestrini,
\npb {568}{2000}{120}, \hepph{9906286}.

\bibitem{Mikolaj} K.~Chetyrkin, M.~Misiak and M.~M\"unz, \plb{400}{1997}{206},
\hepph{9612313}.

\bibitem{counterterm}
M.~Ciuchini, E.~Franco, G.~Martinelli, L.~Reina and L.~
Silvestrini, \plb{316}{1993}{127}, \hepph{9307364}; 
\npb{415}{1994}{403}, \hepph{9304257}; G.~Cella, G.~Curci, G.~Ricciardi and A.~
Vicere, \plb{325}{1994}{227}, \hepph{9401254}.

\bibitem{Adel} K.~Adel and Y.-P.~Yao, \prdd{49}{1994}{4945}, \hepph{9308349}.

\bibitem{GH} 
C.~Greub and T.~Hurth, \prdd{56}{1997}{2934}, \hepph{9703349}.

\bibitem{INFRARED}
A.J.~Buras, A.~Kwiatkowski and N.~Pott, \npb{517}{1998}{353},
\hepph{9710336};
\plb{414}{1997}{157}, \hepph{9707482}, E:\ibid{434}{1998}{459}.  

\bibitem{AG91}  A.~Ali and C.~Greub, \zpc{49}{1991}{431}; \plb{259}{1991}{182}.

\bibitem{Pott}  N.~Pott, \prdd{54}{1996}{938}, \hepph{9512252}.

\bibitem{GHW} C.~Greub, T.~Hurth and D.~Wyler, \plb{380}{1996}{385},
  \hepph{9602281}; \prdd{54}{1996}{3350}, \hepph{9603404}.

\bibitem{Ali:1998rr}
A.~Ali, H.~Asatrian and C.~Greub, \plb{429}{1998}{87},
\hepph{9803314}.

\bibitem{Hurth:2001ja}
T.~Hurth and T.~Mannel,
AIP Conf.\ Proc.\  {\bf 602} (2001) 212,
\hepph{0109041}.

\bibitem{Hurth:2001yb}
T.~Hurth and T.~Mannel, \plb{511}{2001}{196},
\hepph{0103331}.

\bibitem{Bobeth:2000}
C.~Bobeth, M.~Misiak and J.~Urban, 
\npb{574}{2000}{291}, 
\hepph{9910220}.

\bibitem{Gambino}
P.~Gambino, M.~Gorbahn and U.~Haisch, \npb{673}{2003}{238}, 
\hepph{0306079}.

\bibitem{AAGW}
H.~H.~Asatryan, H.~M.~Asatrian, C.~Greub and M.~Walker,
\prdd{65}{2000}{074004}, \hepph{0109140}. 

\bibitem{Asa2}
H.~H.~Asatryan, H.~M.~Asatrian, C.~Greub and M.~Walker,
\prdd{66}{2002}{034009}, \hepph{0204341}.

\bibitem{Ghinculov:2003bx}
A.~Ghinculov, T.~Hurth, G.~Isidori and Y.~P.~Yao,
\hepph{0310187}.

\bibitem{Adrian1}
A.~Ghinculov, T.~Hurth, G.~Isidori and Y.~P.~Yao,
\npb{648}{2003}{254},
\hepph{0208088}.

\bibitem{Asatrian:2002va}
H.~M.~Asatrian, K.~Bieri, C.~Greub and A.~Hovhannisyan,
\prdd{66}{2002}{094013},
\hepph{0209006}.

\bibitem{Adrian2}
A.~Ghinculov, T.~Hurth, G.~Isidori and Y.~P.~Yao,
Nucl.\ Phys.\ Proc.\ Suppl.\  {\bf 116} (2003) 284,
\hepph{0211197}.

\bibitem{Asatrian:2003yk}
H.~M.~Asatrian, H.~H.~Asatryan, A.~Hovhannisyan and V.~Poghosyan,
\hepph{0311187}.

\bibitem{hurth_review}
T.~Hurth,
Rev. Mod. Phys. {\bf 75} (2003) 1159, \hepph{0212304}.

\bibitem{Tarasov}
O.~V.~Tarasov, \prdd{54}{1996}{6479}, \hepth{9606018}.

\bibitem{baikov}
P.~A.~Baikov, \plb{474}{2000}{385}, \hepph{9912421}.

\bibitem{Misiak:1993bc}
M.~Misiak, \npb{393}{1993}{23}, \npb{439}{1995}{461} (E).

\bibitem{Buchalla:1995vs}
G.~Buchalla, A.~J.~Buras and M.~E.~Lautenbacher,
Rev.\ Mod.\ Phys.\  {\bf 68} (1996) 1125,
\hepph{9512380}.

\bibitem{Buras:1994dj}
A.~J.~Buras and M.~Munz,
\prdd{52}{1995}{186},
\hepph{9501281}.

\bibitem{Grinstein:1989}
B.~Grinstein, M.~J.~Savage and M.~B.~Wise, \npb{319}{1989}{271}.

\bibitem{Smirnov}
V.~A.~Smirnov, {\it Renormalization and Asymptotic Expansions} (Birkh\"auser,
Basel, 1991); {\it Applied Asymptotic Expansions in Momenta and Masses}
(Springer-Verlag, Heidelberg, 2001); Mod.\ Phys.\ Lett.\ A {\bf 10} (1995)
1485, \hepth{9412063}.

\bibitem{sunset}
O.~V.~Tarasov, \npb{562}{1997}{455}, \hepph{9703319}.

\bibitem{letter}
H.~H.~Asatryan, H.~M.~Asatrian, C.~Greub and M.~Walker, \plb{507}{2001}{162},
\hepph{0103087}. 

\bibitem{Nir}
Y.~Nir, \plb{221}{1989}{184}.

\bibitem{Ostermaier}
A.~J.~Buras, M.~E.~Lautenbacher and G.~Ostermaier,
\prdd{50}{1994}{3433}, \hepph{9403384}.

\bibitem{Ali:1998sf}
A.~Ali and G.~Hiller,
Eur.\ Phys.\ J.\ C {\bf 8} (1999) 619, \hepph{9812267}.

\bibitem{Beneke:2001at}
M.~Beneke, T.~Feldmann and D.~Seidel,
\npb{612}{2001}{25}, \hepph{0106067}.

\end{thebibliography}
\end{document}